\documentclass{aims}
\usepackage{amsmath}
  \usepackage{paralist}
  \usepackage{graphics} %% add this and next lines if pictures should be in esp format
\usepackage{graphicx}  \usepackage{epstopdf}%This is to transfer .eps figure to .pdf figure; please compile your paper using PDFLeTex or PDFTeXify.
 \usepackage[colorlinks=true]{hyperref}
\hypersetup{urlcolor=blue, citecolor=red}

\def\currentvolume{34}
 \def\currentissue{8}
  \def\currentyear{2014}
   \def\currentmonth{August}
    \def\ppages{3061-3093}
     \def\DOI{10.3934/dcds.2014.34.3061}

\theoremstyle{definition}

\title[On  the nature of large and rogue waves.]{ On  the nature of large and rogue waves.}

\author[Mikhail Kovalyov]{}


 \email{mkovalyo@gmail.com}

\begin{document}
\maketitle

% Enter the first author's name and address:
\centerline{\scshape Mikhail Kovalyov }
\medskip
{\footnotesize
% please put the address of the first author
 \centerline{Department of Mathematics, Sungkyunkwan University,   }
   \centerline{ Suwon, Gyeonggi-do, S. Korea}
} % Do not forget to end the {\footnotesize by the sign }

\bigskip

% The name of the associate editor will be entered by an editorial staff
% "Communicated by the associate editor name" is not needed for special issue.
 \centerline{(Communicated by the associate editor name)}

%The abstract of your paper
\begin{abstract}
In this paper we discuss a model of large and rogue waves in non-necessarily shallow water. We assume that the
relevant portion of the flow is restricted to a near-surface layer,
assumption which enables us    to use the Kadomtsev-Petviashvili equation. The shape and behavior of several types of waves predicted by some singular solutions of the Kadomtsev-Petviashvili equation is compared to the physical waves observed in the ocean.
\end{abstract}

\section{A very short introduction into the history of rogue waves.}

The goal of this paper is to show  that the Kadomtsev-Petviashvili equation  provides a rough  description and classification of large waves and   predicts the existence of so-called "rogue waves".

 Rogue waves   (also known as monster waves, freak waves,  killer waves, extreme waves,  abnormal waves, etc.) are  large ocean surface waves that occur far out in sea and appear spontaneously supposedly out of nowhere, even in calm seas. One such rogue wave caused the Ocean Ranger, then the world's largest offshore platform, to capsize in 1982.      Rogue waves could appear in
any place of the world ocean,    \cite{Divinsky04, Earle75, Gutshabash86, Irvine88, Mori02, Mallory74}.
They present a threat even to large ships and ocean liners.
 Until recently rogue waves were mostly known to exist only through anecdotal evidence provided by those who had encountered them at sea;
 mariners' testimonies  of rogue waves were often treated with disbelief  despite the damages inflicted by such waves on ships. It all changed after  the  scientific measurement  positively confirmed the existence of  a large wave at the Draupner platform, in the North Sea on January 1, 1995.  Since then numerous accounts of rogue waves have been reported, mostly in media.  It's not just ships and offshore structures that need to worry about rogue waves; the U.S. Navy has expressed concern that some Coast Guard rescue helicopters lost at sea may have been struck by rogue waves  \cite{Freeze06}. Can rogue wave  shoot up so high as to present a danger to helicopters?

  It has been observed that large/rogue waves often appear in sets of more than one, the famous sets of threes known as "three sisters" are often quoted. According to some accounts three sisters have been observed in  Lake Superior.
 As unbelievable as the stories of rogue waves might have been, even less believable are the stories of
 the  phenomenon known among Russian sailors as "devyatiy val" (which may be translated as "the ninth super-wave" or "the ninth wall of water") and popularized quite unscientifically by Ivan Aivazovsky in his famous painting of the same name.  According to the folklore, rogue  waves may appear in groups among which the ninth wave is the strongest and the most devastating, hence the name. Ancient Greeks held similar beliefs only it was the tenth rather then the ninth  wave  that was the most powerful.  Is the "devyatiy val" just a fiction   or may  there be a grain of truth in it?

 Rogue waves have now been hypothesized as a cause for many unexplained sea losses over the years including the losses over the Bermuda triangle.  But can  this explanation account for the loss of airplanes?  Although not very likely, the aforementioned Cost Guard report certainly leaves the possibility open.  As far-fetched as it   might be, it is more realistic than the explanations involving "worm-holes", extra-terrestrial life-forms or the paranormal.

There seems to be a fad nowadays to refer to any large oceanic/lake wave as a  "rogue wave". Given the abundance of atmospheric, oceanographic  and geological phenomena on Earth it is very unlikely that all large waves  have the same cause(s) and  exhibit the same behavior.
There are currently many models claiming to describe rogue waves, at least some must be fairly successful in describing some type(s) of large waves.   Since large waves considerably differ from each other in shape and behavior it is very unlikely that one single model can describe all of them.
In this article  the term  "rogue waves" is restricted to the large waves which  appear seemingly out of nowhere in relatively calm waters, have   a relatively short life-span  and are localized in a relatively small region of space.

We will show that  singular solutions of the  Kadomtsev-Petviashvili   equation provide a surprisingly good  description of  the shape of at least some  large non-rogue waves  and lead  to the prediction of the existence of true rogue waves.  We will not address the cause of any of these  waves. The Kadomtsev-Petviashvili   equation is chosen as the simplest nonlinear equation of water waves in $\textbf{R}^3$ many of whose solutions may be computed explicitly. One, of course, should understand that  the Kadomtsev-Petviashvili   equation may provide   only a limited description of the phenomena. It is quite surprising that the  Kadomtsev-Petviashvili   equation turns out to be so useful as both the singular solutions and the large waves exit on the borderline of applicability  of the Kadomtsev-Petviashvili  equation  derived to describe waves in shallow water. We will address the applicability of the Kadomtsev-Petviashvili  equation in the last section.

\section{ Derivation of the Kadomtsev-Petviashvili   equation.   }

The original derivation of the Kadomtsev-Petviashvili   equation was reported in \cite{kp}, since then different derivations based on essentially the same ideas appeared in  numerous publications.  For the completeness's sake and because some parts of the derivation will be referred to later on in the paper, we shall reproduce    the derivation  here. Typically the Kadomtsev-Petviashvili   equation is derived for shallow water waves; however, as   point 4 of the derivation below shows the validity of the Kadomtsev-Petviashvili   equation may be extended to waves in non-shallow  water with the vertical component
of the motion basically varying only in a near-surface layer.
 To do so consider the motion of fluid, subject to the following assumptions:

\vskip3mm \noindent 1.  The fluid is considered in the region of the three-dimensional space $\textbf{R}^3$ described by
$$ (x^\prime,y^\prime)\in \Omega\subset  \textbf{R}^2,   -h<z^\prime<\epsilon h \eta(x^\prime,y^\prime,t^\prime),\eqno{\rm
(2.1)}
$$
where $h=  \textmd{constant} >0,$ and $ \epsilon= \textmd{constant} >0 $ is sufficiently small.  The set $(x^\prime,y^\prime)~\in~ \Omega,$ $  z^\prime=-h $ is  the bed of the fluid and the function  $z^\prime=\epsilon h \eta(x^\prime,y^\prime,t^\prime)$ represents the elevation of the free surface above the reference plane $z^\prime=0$ at time $t^\prime.$  We shall refer to   variables $x^\prime, y^\prime, z^\prime, t^\prime$ as {\it physical coordinates },  to $\epsilon h \eta(x^\prime,y^\prime, t^\prime)$ as {\it physical elevation } and to $x^\prime, y^\prime,  t^\prime, z^\prime= \epsilon h \eta$  as {\it physical variables}.

\vskip3mm \noindent 2.  The  fluid is   irrotational, i. e. the velocity vector $\boldsymbol{u^\prime }(x^\prime,y^\prime,z^\prime,t^\prime)$    satisfies $\nabla \times \boldsymbol{u^\prime} = 0 $ for all $t^\prime$. Consequently the fluid has a velocity potential $\phi$, i.e  $$\boldsymbol{u^\prime}=\epsilon\nabla \phi. \eqno{\rm
(2.2)}
$$
The assumption of irrotational flow means that there are no non-uniform underlying currents.
\vskip3mm\noindent 3. The  fluid is   incompressible, i. e. the velocity vector $\boldsymbol{u^\prime}$  and velocity potential $\phi$   satisfy
$$
\nabla \cdot \boldsymbol{u^\prime} =  \epsilon\left[\frac{\partial^2 \phi}{\partial x^{\prime \, 2}}+\frac{\partial^2 \phi}{\partial y^{\prime \, 2}}+\frac{\partial^2 \phi }{\partial z^{\prime \,2}}\right]=0 \hskip2mm   \textmd{in the region }  -h<z^\prime<\epsilon h \eta(x^\prime,y\prime,t^\prime).{\rm
(2.3)}
$$
 \vskip3mm \noindent 4.  Typically the derivation of the Kadomtsev-Petviashvili   equation  is based on the assumption that the  bed is impermeable, i.e the velocity and potential  satisfy
$$
\boldsymbol{u^\prime}\cdot \boldsymbol{n} =\epsilon\frac{\partial \phi}{\partial z^\prime}= 0\hskip15mm \mbox{ on }  z^\prime = -h.
\eqno{\rm
(2.4)}
$$
However, equation (2.4) is also valid if we assume that the flow is horizontal below the plane  $z^\prime= -h $, i.e $\boldsymbol{u^\prime}\cdot \boldsymbol{n}=0, $ for $z^\prime \leq  -h.$ This trivial assumption allows us to extend the derivation to flows in non-shallow water.

\vskip3mm \noindent 5.   On the free surface $z^\prime=\epsilon h \eta(x^\prime,y^\prime,t^\prime)$ we require that  the fluid  flows  along the free surface   $z^\prime=\epsilon h \eta(x^\prime,y^\prime,t^\prime)$ without ever leaving the surface, i. e.   the vertical component of the velocity satisfies
 $$
 \displaystyle\frac{\partial \phi}{\partial z^\prime}\hskip-1mm =\hskip-1mm h\left[\frac{\partial \eta}{\partial t^\prime}+ \boldsymbol{u}\cdot\nabla \eta \right]\hskip-1mm =\hskip-1mm h\left[\displaystyle\frac{\partial \eta}{\partial t^\prime} + \epsilon\displaystyle\frac{\partial \phi}{\partial x^\prime} \displaystyle\frac{\partial \eta}{\partial x^\prime} +\epsilon \displaystyle\frac{\partial \phi}{\partial y^\prime }\displaystyle\frac{\partial \eta}{\partial y^\prime}\right]   \mbox{ on  } z^\prime \hskip-1mm =\hskip-1mm\epsilon h\eta(x^\prime,y^\prime,t^\prime).
\eqno{\rm
(2.5)}
$$

\vskip3mm\noindent 6.  We assume that the density of the fluid is  constant, i. e.
$$
\rho= \textmd{constant}.
\eqno{\rm
(2.6)}
$$

\vskip3mm \noindent 7. The fluid is  inviscid, i. e. there is no frictions and the only forces present are gravity and pressure.
On the free surface   $z^\prime=\epsilon h \eta(x^\prime,y^\prime,t^\prime)$   the  pressure  is given
$
p= p_a - s\nabla\cdot\boldsymbol{n}=p_a-s\left(\displaystyle\frac{1}{R_1}+ \displaystyle\frac{1}{R_2}\right)
$
where  $p_a$ is atmospheric pressure, $s$ is surface tension, $\boldsymbol{n}$ is the unit normal to $z^\prime=\epsilon h \eta(x^\prime,y^\prime,t^\prime),$     $ \displaystyle\frac{1}{R_1}=\displaystyle\frac{\epsilon h \eta_{x^\prime x^\prime}(1+\epsilon^2 h^2\eta_{y^\prime}^2)}{\big(1+\epsilon^2h^2\eta_{x^\prime}^2+\epsilon^2h^2\eta_{y^\prime}^2\big)^{3/2}} $ and $ \displaystyle\frac{1}{R_2}=\displaystyle\frac{\epsilon h \eta_{y^\prime y^\prime}(1+\epsilon^2 h^2\eta_{x^\prime}^2)}{\big(1+\epsilon^2h^2\eta_{x^\prime}^2+\epsilon^2h^2\eta_{y^\prime}^2\big)^{3/2}}$ are the reciprocals of the curvature radii. We may assume $p_a=0 $ and use assumptions (2.6)  to simplify  the last two expressions to $ \displaystyle\frac{1}{R_1}= \epsilon h\eta_{x^\prime x^\prime}$ and $ \displaystyle\frac{1}{R_2}=\epsilon h \eta_{y^\prime y^\prime}$ which in turn simplifies the equation for $p$ to
$$
p \approx -s\epsilon h(\eta_{x^\prime x^\prime}+\eta_{y^\prime y^\prime})  \mbox{ at }  z=\epsilon h\eta.
\eqno{\rm
(2.7)}
$$

\noindent 8. The potential $\phi$ satisfies the Bernoulli's equation $ \epsilon\displaystyle\frac{\partial \phi}{\partial t^\prime}+\epsilon^2\displaystyle\frac 12 \left \vert \nabla \phi\right\vert^2- \displaystyle\frac{ s}{\rho}  \epsilon h(\eta_{x^\prime x^\prime}+\eta_{y^\prime y^\prime})+ gz^\prime= C(t^\prime) $, where $C(t^\prime)$  is an arbitrary function.  Allowing  $\displaystyle\frac{\partial \phi}{\partial t^\prime}$ to absorb the term $  -C(t)$ we may rewrite the restriction of the Bernoulli's equation to the free surface $z^\prime=\epsilon h\eta(x^\prime,y^\prime,t^\prime)$ as
$$
\displaystyle\frac{\partial \phi}{\partial t^\prime}+\displaystyle\frac 12 \epsilon\left \vert \nabla \phi\right\vert^2- \displaystyle\frac{ s}{\rho}   h\big(\eta_{x^\prime x^\prime}+\eta_{y^\prime y^\prime}\big)+    h g \eta=0   \hskip15mm \mbox{ on } z^\prime = \epsilon h\eta(x^\prime,y^\prime,t^\prime).
\eqno{\rm
(2.8)}
$$

\vskip3mm \noindent 9. The fluid  motion is  essentially one-dimensional  with the  main motion  occurring along the $x^\prime- \textmd{axis}$, the motion and variations  along the $y^\prime-\textmd{axis}$ are smaller than the motion and variations along the  $x^\prime-\textmd{axis,} $  we symbolically write it as
$$
\displaystyle\frac{\partial  }{\partial y^\prime} =  o\left( \displaystyle\frac{\partial  }{\partial x^\prime}\right)
\eqno{\rm
(2.9)}
$$

For $ \epsilon$  sufficiently small the first approximation is obtained by setting $\epsilon=0$. Equations   (2.3),  (2.4), (2.5), (2.8)  become
$$
\nabla^2 \phi = 0 \hspace{3.5cm} -h<z^\prime<\eta(x^\prime,y^\prime,t^\prime)
\eqno{\rm
(2.10)}$$
$$
\frac{\partial \phi}{\partial z^\prime}= 0 \hspace{5.2cm} \mbox{at} \hspace{2mm}z^\prime = -h
\eqno{\rm
(2.11)}
$$
$$
\frac{\partial\phi}{\partial z^\prime}= \frac{\partial\eta}{\partial t^\prime} \hspace{5.2cm}\mbox{at}\hspace{2mm} z^\prime = 0
\eqno{\rm
(2.12)}
$$
$$
\frac{\partial \phi}{\partial t^\prime}+ g\eta - \displaystyle\frac{s}{\rho} (\eta_{x^\prime x^\prime}+\eta_{y^\prime y^\prime})=0\hspace{2.0cm}\mbox{at}\hspace{2mm} z^\prime=0
\eqno{\rm
(2.13)}
$$

We may look for a solution of system   (2.10) -  (2.13) in the form
$$
\eta(x^\prime,y^\prime,t^\prime)= e^{i (kx^\prime+ly^\prime-\omega t^\prime)},
\eqno{\rm
(2.14)}
$$
and
$$
\phi(x^\prime,y^\prime,z,t^\prime)=f_{kl}(z)e^{i(kx^\prime+ly^\prime-\omega t^\prime)}.
\eqno{\rm
(2.15)}
$$
Substituting (2.15) into  (2.10), we obtain   equation
$
f_{kl}''-(k^2+l^2)f_{kl}=0
$
whose general solution is
$
f_{kl}(z)=C_1 e^{\sqrt{k^2+l^2}\;  z}+ C_2 e^{-\sqrt{k^2+l^2}\;  z},
$
which upon substitution into  (2.15)   gives us
$$
\phi(x^\prime,y^\prime,z,t^\prime)=\left(C_1 e^{\sqrt{k^2+l^2}\; z}+ C_2 e^{-\sqrt{k^2+l^2}\;  z}\right)e^{i(kx^\prime+ly^\prime-\omega t^\prime)}.
\eqno{\rm
(2.16)}
$$
Here $C_1$ and $ C_2  $ are constants of integration which can be determined from
boundary conditions (2.11) and (2.12)
$$
\hskip-1mm C_1=- \frac{i \omega}{\sqrt{k^2+l^2}\; \Big(1-e^{-2h\sqrt{k^2+l^2}\; }\Big)}, \;
C_2=- \frac{i \omega e^{-2h\sqrt{k^2+l^2}  }}{\sqrt{k^2+l^2}  \Big(1-e^{-2h \sqrt{k^2+l^2}  }\Big)}
\eqno{\rm
(2.17)}
$$
Substituting (2.17) into (2.16) we obtain
$$
\phi(x^\prime,y^\prime,z,t^\prime)=- \frac{i\omega}{\sqrt{k^2+l^2}\;}\frac{\cosh\Big[ (z+h)\sqrt{k^2+l^2}\Big]}{\sinh\Big(h \sqrt{k^2+l^2}\Big)}e^{i(kx^\prime+ly^\prime-\omega t^\prime)}
\eqno{\rm
(2.18)}
$$
Boundary condition (2.13) gives us the relationship between $\omega$ and $\sqrt{k^2+l^2}\;$ in the form
$$
\omega^2=\sqrt{k^2+l^2}\;\left[g+
\displaystyle\frac s\rho(k^2+l^2)\right]\tanh\left( h\sqrt{k^2+l^2}\right)
\eqno{\rm
(2.19)}
$$
Solving (2.19) for positive$\omega $ and using Taylor approximation along with $l\ll k$ which is due to   (2.9), we obtain
$$
\hskip-4.5cm \omega=\sqrt{\sqrt{k^2+l^2}\;\left[g+\displaystyle\frac s\rho (k^2+l^2)\right]\tanh(h\sqrt{k^2+l^2}\;)}=
$$
$$
 \sqrt{gh}   \left[k+\displaystyle\frac{l^2}{2k}+\displaystyle\frac{h^2}{6}\left(1-\displaystyle\frac{3s}{g\rho h^2} \right)k^3\right]+  \textmd{lower order terms}
\eqno{\rm
(2.20)}
$$
As the first approximation to $\eta$ we may consider
$$
\eta=e^{i\left\{kx^\prime+ly^\prime-\sqrt{gh}   \left[k+\frac{l^2}{2k}+\frac{h^2}{6}\left( \frac{3s}{g\rho h^2} -1 \right)k^3\right]\; t^\prime\right\}}+o(k^3)= e^{ik(x^\prime-\sqrt{gh }\; t^\prime)}+o(k ),
\eqno{\rm
(2.21)}
$$
 whose first term satisfies
$$
\displaystyle\frac{\partial}{\partial x^\prime}\left[\displaystyle\frac 1{\sqrt{gh}}  \eta_{ t^\prime} +\eta_{ x^\prime} +  \displaystyle\frac{g\rho h^2-3s}{6g\rho  }  \eta_{ x^\prime x^\prime x^\prime}\right]+  \displaystyle\frac{1}{2}\eta_{ y^\prime y^\prime}=0
\eqno{\rm
(2.22)}$$
 known as {\it linear  Kadomtsev-Petviashvili equation. }The nonlinear {\it physical   Kadomtsev-Petviashvili equation} is obtained by adding to the left hand-side a nonlinear term proportional to $\displaystyle\frac{\partial\eta ^2 }{\partial x} $  based on some physical argument. The derivation of (2.22) is based on the premise that  waves are   of the form (2.14)-(2.15) and  $k$, $l$ and $h\sqrt{k^2+l^2}$ are small, i.e.   waves (2.14)-(2.15) have long wave length considerably exceeding  the depth. Due to this  the Kadomtsev-Petviashvili  equation is often said to describe   long waves in shallow water.

 The derivation above is based on the assumption that the solutions of the Kadomtsev-Petviashvili  equation are of the form (2.14) - (2.15) which is quite restrictive.   We may somewhat generalize the derivation using multi-scale approach by  introducing  dimensionless coordinates
$$
T\hskip-1mm=\hskip-1mm t^\prime\sqrt{\displaystyle\frac{\epsilon^{3 }g }{h }},\;  X\hskip-1mm=\hskip-1mm\sqrt{\epsilon}\displaystyle\frac{x^\prime- t^\prime \sqrt{g  h}}{h } ,\;  Y\hskip-1mm=\hskip-1mm {\epsilon} \displaystyle\frac{y^\prime}{h },\;  Z\hskip-1mm=\hskip-1mm\displaystyle\frac{z^\prime}{h},\;   \Phi\hskip-1mm=\hskip-1mm \phi\sqrt{\displaystyle\frac{\epsilon }{gh^3}},\;  S=\displaystyle\frac{s}{\rho h^2 g},\eqno{\rm
(2.23)}
$$
 which transform equations (2.3),  (2.4), (2.5), (2.8)  into
$$
  \epsilon  \frac{\partial^2 \Phi}{\partial X^2}+   \epsilon^2   \frac{\partial^2 \Phi}{\partial Y^2}+\frac{\partial^2 \Phi }{\partial Z^2}=0 \hskip5mm  \textmd{in the region}   -1<Z<  \epsilon \eta ,
\eqno{\rm
(2.24)}
$$
$$
 \frac{\partial \Phi}{\partial Z}= 0\hskip15mm \mbox{ on }  Z = -1,
\eqno{\rm
(2.25)}
$$
$$
 \displaystyle\frac{\partial \Phi}{\partial Z}  =-   \epsilon\displaystyle\frac{\partial \eta}{\partial X} + \epsilon^2    \displaystyle\frac{\partial \eta}{\partial T} + \epsilon^2 \displaystyle\frac{\partial \Phi}{\partial X} \displaystyle\frac{\partial \eta}{\partial X} + \epsilon^3 \displaystyle\frac{\partial \Phi}{\partial Y }\displaystyle\frac{\partial \eta}{\partial Y} \hskip10mm   \mbox{ on  } Z = \epsilon  \eta ,
\eqno{\rm
(2.26)}
$$
$$
   \eta -S \epsilon\eta_{XX}-S\epsilon^2 \eta_{YY}  -  \displaystyle\frac{\partial \Phi}{\partial X}+\epsilon  \displaystyle\frac{\partial \Phi}{\partial T}+\displaystyle\frac 12    \left(\displaystyle\frac{\partial \Phi}{\partial Z}\right)^2\hskip-2mm
 +\displaystyle\frac 12 \epsilon   \left(\displaystyle\frac{\partial \Phi}{\partial X}\right)^2\hskip-2mm+\displaystyle\frac 12 \epsilon^2  \left(\displaystyle\frac{\partial \Phi}{\partial Y}\right)^2\hskip-2mm=0     \mbox{ on }  Z = \epsilon  \eta .
\eqno{\rm
(2.27)}
$$

We look for solutions of system (2.24)-(2.27)  in the form
$$
\Phi=\Phi_0+\epsilon\Phi_1+\epsilon ^2 \Phi_2+\dots, \;\;\eta=\eta_0+\epsilon\eta_1+ \dots
\eqno{\rm
(2.28)}
$$
Substituting (2.28)  into (2.24)-(2.25)  we obtain
$$
 \Phi_{0 ZZ}+\epsilon \Big(\Phi_{0 XX}+\Phi_{1 ZZ}\Big)+\epsilon^2\Big(
\Phi_{1 XX}+\Phi_{0 YY}+ \Phi_{2 ZZ} \Big)+O(\epsilon^3)=0,
\eqno{\rm
(2.29)}
$$
$$
 \Phi_{0  Z}=   \Phi_{1 Z}=   \Phi_{2 Z}=0\hskip15mm \mbox{ on }  Z = -1,
\eqno{\rm
(2.30)}
$$
$$
\Phi_{0Z}+\epsilon\Big(\Phi_{1Z}+ \eta_{0X}\Big)+\epsilon^2\Big(\Phi_{2Z}+ \eta_{1X}-\eta_{0T}-
\Phi_{0X}\eta_{0X} \Big)+O(\epsilon^3)=0,  \mbox{ on }  Z = \epsilon\eta,
\eqno{\rm
(2.31)}
$$
 $$
\eta_0 +\displaystyle\frac12 \Phi_{0Z}^2-\Phi_{0X} +\epsilon\Big( \eta_1-S \eta_{0XX}  -\Phi_{1X} +\Phi_{0T}+\Phi_{0Z}\Phi_{1Z}+\displaystyle\frac12 \Phi_{0X}^2\Big) +O(\epsilon^2)=0,  \mbox{ on }  Z = \epsilon\eta.
 \eqno{\rm
(2.32)}
$$
Substituting  (2.28) into (2.29) - (2.30)  and equating to zero the coefficients of $\epsilon^0, \epsilon^1, \epsilon^2 $  we obtain
$$\begin{array}{ll}
 \Phi_0=A_0(X,Y,T),\;\Phi_1=-\displaystyle\frac 12  A_{0XX} (Z+1)^2+A_1(X,Y,T)
 ,\\  \Phi_2= \displaystyle\frac 1{24} A_{0XXXX} (Z+1)^4- \displaystyle\frac 1{2}  \Big[A_{0YY}+A_{1XX}\Big] (Z+1)^2+A_2(X,Y,T).
 \end{array}\eqno{\rm
(2.33)}$$
Equating to zero the coefficient of  $\epsilon^0$ in  (2.32)   and using $ \Phi_{0Z}=A_{0Z}=0$ we obtain
$$
\eta_0 =   \Phi_{0X}= A_{0X}.\eqno{\rm
(2.34)}$$
Equating the coefficient $\eta_1-S \eta_{0XX}   -\Phi_{1X}+\Phi_{0T} +\displaystyle\frac 12 \Phi_{0X}^2 $ of  $\epsilon$ in (2.32)   to zero  and using   $ \Phi_{0Z}=A_{0Z}=0$, $Z= \epsilon \eta$  as well as (2.34)  we obtain
$$
\eta_1-S \eta_{0XX}   -A_{1X}+\displaystyle\frac 12\eta_{0XX}+A_{0T} +\displaystyle\frac 12 \eta_{0}^2=0.\eqno{\rm
(2.35)}
$$
Substituting $ \Phi_{0Z}=A_{0Z}=0$, $Z= \epsilon \eta$, $ \Phi_{1Z}=-   A_{0XX} $, $\Phi_{2Z}= \displaystyle\frac 1{6}  A_{0XXXX}  -   A_{0YY} -   A_{1XX} $ into equation (2.31) we obtain that the coefficients of  $\epsilon^0$ and   $\epsilon$ trivially vanish while equating to zero the coefficient $- A_{0XX}\eta_0+   \Phi_{2Z}+  \eta_{1X}-\eta_{0T}-
\Phi_{0X}\eta_{0X} $ of     $\epsilon^2$ gives us
$$
 -2\eta_{0X}\eta_0+ \displaystyle\frac 1{6}  \eta_{0XXX}  -   A_{0YY}- A_{1XX}  + \eta_{1X}-\eta_{0T} =0.
\eqno{\rm (2.36)}
$$
Differentiation of  (2.35) in $X$ with the help of (2.34) further gives us
$$
\eta_{1X}-S \eta_{0XXX}   -A_{1XX}+\displaystyle\frac 12\eta_{0XXX}+\eta_{0T} + \eta_{0} \eta_{0X} =0. \eqno{\rm
(2.37)}
$$
Subtracting  (2.36)  from  (2.37)  we obtain
$$
2 \eta_{0T}+\left(\displaystyle\frac 13-S\right)\eta_{0XXX}+3\eta_{0 }\eta_{0X}+ A_{0YY} =0 . \eqno{\rm
(2.38)}
$$
Differentiation of  (2.38) in $X$ yields, with the help of (2.34),
$$
\displaystyle\frac{\partial}{\partial X}\left[2 \eta_{0T}+\left(\displaystyle\frac 13-S\right)\eta_{0XXX}+3\eta_{0 }\eta_{0X}\right]+\eta_{0YY} =0, \eqno{\rm
(2.39)}
$$
or in dimensional coordinates
$$
\displaystyle\frac{\partial}{\partial x^\prime}\left[ \displaystyle\frac{1}{  \sqrt{  g h} } \;\eta_{0t^\prime}+  \eta_{0x^\prime}+ \displaystyle\frac{\rho g h^2-3s}{6\rho g  }\eta_{0x^\prime x^\prime x^\prime}+\displaystyle\frac {3\epsilon }{2}\eta_{0 }\eta_{0x^\prime}\right]+\displaystyle\frac{1 }{2 }\eta_{0y^\prime y^\prime} =0, \eqno{\rm
(2.40)}
$$
which we shall call {\it physical Kadomtsev-Petviashvili   equation. }
Notice that even if we dropped  $\eta_{YY}$ in (2.27), or, equivalently,  assumed that $R_2=+\infty$ in the derivation of (2.7), $\eta_{0YY}$  would still appear in (2.39).

 The change of variables
$$  \alpha^2=\displaystyle\frac{2\rho g  }{\rho g h^2-3s},    \; t^\prime=  \displaystyle\frac{3\alpha^2\; t }{\sqrt{gh}} ,   x^\prime=   x +{3\alpha^2} t ,\;y^\prime=\frac {y }{\sqrt{2}\thinspace} , \; \eta_0=\displaystyle\frac {4f}{3\epsilon\alpha^2}  ,
\eqno{\rm
(2.41a)}
$$
or equivalently,
$$ \alpha^2=\displaystyle\frac{2\rho g  }{\rho g h^2-3s},    \;   t  =\displaystyle\frac{\sqrt{gh} \; t^\prime}{3\alpha^2  },    x = x^\prime-{\sqrt{gh} \; t^\prime},\;   {y }=y^\prime{\sqrt{2}\thinspace} , \; f=\displaystyle\frac {3\epsilon\alpha^2} 4 \eta_0,
\eqno{\rm
(2.41b)}
$$
turns   equation (2.40) into
$$
\displaystyle\frac{\partial}{\partial x}\Big[ f_t+f_{xxx}+6ff_x\Big]+3\alpha^2 f_{yy}=0,   \quad \alpha^2=\pm 1,
\eqno{\rm
(2.42)}
$$
which is known as {\it the Kadomtsev-Petviashvili   equation} or simply KP.
Notice that the physical  Kadomtsev-Petviashvili   equation given by (2.40) differs from the  Kadom- tsev-Petviashvili   equation given by (2.42). It is the physical  Kadomtsev-Petviashvili   equation (2.40) that describes the physical elevation $\epsilon h \eta(x^\prime,y^\prime,t^\prime) $ in terms of physical coordinates  $x^\prime,y^\prime,t^\prime $; equation (2.42) describes a nonphysical quantity  $f(x,y,t)$  in terms of  variables $x, y, t$ in  the frame of reference moving with velocity $\sqrt{gh}$  in the physical frame of reference $x^\prime,y^\prime,t^\prime $. The quantity $\sqrt{gh}$ is the shallow water speed for irrotational travelling waves
propagating in water of mean depth $h$, see  \cite{Johnson97}

Notice that formulas (2.2) and  (2.34) imply that for $z^\prime\geq -h$
$$
  u^\prime_{x^\prime} =\epsilon \displaystyle\frac{\partial \phi }{\partial x^\prime} =\epsilon\sqrt{gh}\;\Phi_{0X}+o(\epsilon )
=\epsilon\sqrt{gh}\;\eta_0+o(\epsilon ), \eqno{\rm
(2.43)}
$$
where,  according to  (2.2),  $u^\prime_{x^\prime} $ is just the $x^\prime-$component of the velocity of the flow.   Thus  the physical Kadomtsev-Petviashvili   equation   may be viewed as an equation not only for   elevation $\eta_0$ but also for the   $x^\prime-$component of the velocity of the flow $u^\prime_{x^\prime}$.   Moreover, the first two formulas in (2.33) and formula (2.34) imply that for $z^\prime\geq -h$
$$u^\prime_{z^\prime}= \epsilon\displaystyle\frac{\partial \phi }{\partial z^\prime}=\sqrt{gh\epsilon^3\;} \;\Phi_{1Z}+o\left(\epsilon^{3/2}  \right)=-\sqrt{gh\epsilon^3}\;\eta_{0X}+\left(\epsilon^{3/2}  \right) =-  \epsilon (h+z^\prime)\sqrt{gh}  \;\displaystyle\frac{\partial \eta_0}{\partial x^\prime}+o\left(\epsilon^{3/2}  \right),\eqno{\rm(2.44)}$$
keeping in mind that $\displaystyle\frac{\partial \eta_0}{\partial x^\prime}=O(\sqrt{\epsilon})$ due to (2.23).

Formulas (2.43), (2.44) give us the components of the velocity of motion of the fluid  not   of the  profile $\epsilon h \eta(x^\prime,y^\prime,t^\prime) $.  Whereas the profile $\epsilon h \eta $ may move in one direction the fluid on the surface $z^\prime=\epsilon h \eta(x^\prime,y^\prime,t^\prime) $ may move in another direction.

\section{ Special solutions of KP and corresponding physical waves.    }
The simplest solution of KP is a soliton wall solution
$$
f(x,y,t)=2\frac{\partial^2}{\partial x^2}\ln \left(1
+\frac{c   \enskip e^{(p +q )x+(q ^2-p ^2)  y-(p+q)\big[(p +q)^2+3\alpha^2(p-q)^2\big]t}}{p +q }\right) ,
\eqno{\rm (3.1)}
$$
where $c, \;p, \; q  $ are  real  constants. Nonlinear superposition of $N$ soliton walls is given by
$$
f(x,y,t)=2\frac{\partial^2}{\partial x^2}\ln \det{\pmb  A},
\eqno{\rm (3.2a)}
$$
where
$$
{\pmb  A}_{mn}=\delta_{mn}
+\frac{c_n  \enskip e^{(p_n+q_n)x+(q_n^2-p_n^2)  y-(p_n+q_n)\big[(p_n +q_n)^2+3\alpha^2 (p_n-q_n)^2\big]t}}{p_n+q_m},~~~~m,
n=1\dots N,\eqno{\rm (3.2b)}
$$
$\delta_{mn}  $ are the regular Kronecker symbols, $c_n,\; p_n,\; q_n p $ are     real constants.

We may determine the velocity  $\pmb v=(v_x, v_y)$ of the motion of a single soliton wall   by setting $\displaystyle\frac{\textmd{d}}{\textmd{d}t}\Big[ {(p +q )x+(q ^2-p ^2)  y-(p+q)\big[(p +q)^2+3\alpha^2(p-q)^2\big]t}\Big]=\newline  (p +q )v_x+(q ^2-p ^2)  v_y -4(p ^3+q ^3)  =0,$ which gives us

$ $
$$  \displaystyle\frac{1}{ (p +q)^2+3\alpha^2(p-q)^2  }v_x + \displaystyle\frac{  q-p  }{ (p +q)^2+3\alpha^2(p-q)^2  }v_y = 1. \eqno{\rm (3.3)}
$$
So velocity is not defined uniquely but   up to an additive term proportional to the unit vector
$\bigg(\displaystyle\frac{p -q }{\sqrt{1+(p -q )^2}},  \displaystyle\frac{1 }{\sqrt{1+(p-q )^2}}  \bigg)$.

The components $\pmb v=(v_x, v_y)$ of the velocity  of motion of the superposition of two soliton walls satisfy the system of equations
$$\begin{array}{ll}
   v_x +  (q_1 -p_1)   v_y =  (p_1 +q_1)^2+3\alpha^2(p_1-q_1)^2 \\\\
 v_x +  (q_2 -p_2)   v_y = (p_2 +q_2)^2+3\alpha^2(p_2-q_2)^2
\end{array}\eqno{\rm (3.4)}$$
and are uniquely defined provided the determinant of the system is nonzero, i.e. $p_1-q_1\neq p_2-q_2$. We cannot attach a meaningful definition of velocity to the motion of nonlinear superposition of three or more soliton walls.  The velocity of the soliton wall $\pmb v=(v_x, v_y)$ is not to be confused with the velocity of the flow given by the vector $\pmb u^\prime,$ the former is the velocity of motion of the profile while the latter is the velocity of motion of the fluid. The two velocities are not the same. While in a periodic traveling wave with no underlying
current the wave profile moves without change of form in one direction,
the particles beneath the surface wave move backwards and forwards,    \cite{Constantin06}. In contrast to this,
for solitary waves all particles move in the direction of wave propagation,
having an ascending or descending path depending on their position
relative to the wave crest,    \cite{Constantin07}.

Solutions (3.1), (3.2) have been well studied with references too numerous to mention so we shall skip the discussion of (3.1), (3.2).

One may notice that while the derivation (2.14)-(2.22) of the Kadomtsev-Petviashvili equation starts with the assumption that  solutions are of the form (2.14)-(2.15), functions  (2.14)-(2.15) themselves are not solutions of KP  \cite{Kovalyov05}.    In search for solutions of KP of the form close to  (2.14)-(2.15)   they arrived at
$$
\begin{array}{lll}
 &f(x,y,t)=2\displaystyle\frac{\partial^2}{\partial x^2}\ln  \Big[2\lambda\Upsilon   - {\sin 2\Gamma } \Big],&\hskip1mm  {\rm(3.5a)}\\
 &\Upsilon =\rho +x\cos (\alpha\chi) +2\Big[\displaystyle\frac{\lambda\sin (\alpha\chi)}{\alpha}-\mu\cos (\alpha\chi) \Big]y +& \\ & \hskip34mm 12\Big[\lambda^2\cos(\alpha\chi)-\alpha^2\mu^2\cos (\alpha\chi) +2\alpha \lambda\mu\sin (\alpha\chi)\Big]t,
  &  \; {\rm(3.5b)} \\
 &\Gamma=  \gamma+\lambda x-2\lambda\mu y+4\lambda(\lambda^2-3\alpha^2\mu^2)t,
 &\hskip1mm {\rm(3.5c)}
\end{array}
$$
where  $\lambda, \mu, \chi, \gamma, \rho  $ are some  constants. If the constants   $\lambda, \mu, \chi, \gamma, \rho  $ are real so is the solution (3.5).  Nonlinear superposition of such solutions is of the form
$$
u(x,y,t)=2\frac{\partial^2}{\partial x^2}\ln  \det{\pmb  K}
\eqno{{\rm(3.6a)}}
$$
where ${\pmb  K}$ is an $N\times N$ matrix with the entries
$$
{\pmb  K}=\left(
\begin{array}{ccccc}
K_{11}&K_{12} &\dots& K_{1N} \\
K_{21} &K_{22}&\dots& K_{2N} \\
\vdots&\vdots&\vdots&\vdots \\
K_{N1} &K_{N2} &\dots &K_{NN}
\end{array}\right)   \eqno{\rm (3.6b)}
$$
$$
\begin{array}{rll}
&K_{nn}= \Upsilon_n  -\displaystyle\frac{\sin2\Gamma_n}{2\lambda_n},
   &{\rm (3.6c)}                 \\ &&\\
&K_{nk}=   \left[
\displaystyle\frac{(\lambda_n-\lambda_k)\sin(\Gamma_n-\Gamma_k)}{\alpha^2(\mu_n-\mu_k)^2+
(\lambda_n-\lambda_k)^2}- \displaystyle\frac{(\lambda_n+\lambda_k)\sin(\Gamma_n+\Gamma_k)}{\alpha^2 (\mu_n-\mu_k)^2+(\lambda_n+\lambda_k)^2}\right]
+&  \\& &\\
  &\hskip1mm  \alpha     \left[  \displaystyle\frac{(\mu_n-\mu_k)\cos(\Gamma_n+\Gamma_k)}{\alpha^2(\mu_n-\mu_k)^2+(\lambda_n+\lambda_k)^2}
   -
\displaystyle\frac{(\mu_n-\mu_k)\cos(\Gamma_n-\Gamma_k)}{\alpha^2 (\mu_n-\mu_k)^2+(\lambda_n-\lambda_k)^2}\right], \enspace n\ne k
   &{\rm (3.6d)}  \\ &&\\
&\Upsilon_n=\rho_n+x\cos (\alpha\chi_n) +2\Big[\displaystyle\frac{\lambda_n\sin (\alpha\chi_n)}{\alpha}-\mu_n\cos (\alpha\chi_n) \Big]y\enskip+&\\& &\\
&\hskip20mm 12\Big[\lambda_n^2\cos(\alpha\chi_n)-\alpha^2\mu_n^2\cos (\alpha\chi_n) +2\alpha \lambda_n\mu_n\sin (\alpha\chi_n)\Big]t,
 &{\rm(3.6e)} \\ &&\\
&\Gamma_n=  \gamma_n+\lambda_n x-2\lambda_n\mu_n y+4\lambda_n(\lambda_n^2-3\alpha^2\mu_n^2)t,
 &{\rm(3.6f)}
\end{array}
$$
where $\lambda_n \textmd{{'s,}}$$\mu_n\textmd{'s,}$$\chi_n\textmd{'s, }$$\gamma_n\textmd{'s, }$$\rho_n\textmd{'s, }$ are   constants.  Change of constants $\rho_n\to -\rho_n, \, \chi_n\to  \dfrac{\chi_n}{\alpha}+\pi$ allows us to change the sign in front of $\Upsilon_n$ while the change $\gamma_n\to \gamma_n\pm \displaystyle \frac {\pi}{2}$ for all $n=1,2\dots,N$ allows us to change the sign  in front of all   $\displaystyle\frac{(\lambda_n+\lambda_k)\sin(\Gamma_n+\Gamma_k)}{\alpha^2 (\mu_n-\mu_k)^2+(\lambda_n+\lambda_k)^2}, \;\newline  \displaystyle\frac{(\mu_n-\mu_k)\cos(\Gamma_n+\Gamma_k)}{\alpha^2(\mu_n-\mu_k)^2+(\lambda_n+\lambda_k)^2}.$

Solution (3.5)   is singular and   approaches $-\infty $  as we move towards the singular curve $2\lambda\Upsilon -  {\sin  2\Gamma }=0.  $  To be physically plausible it needs to be regularized to remove infinities; since there is no clear way to determine what kind of regularization is most suitable we choose   an ad-hoc regularization
$$
f \to F= \displaystyle\frac{f}{\vert f\vert }\ln \Big[\ln \Big(\Big \vert e^f-1 \Big\vert+1\Big)+1\Big] \eqno{\rm(3.7a)}
$$
due to its properties
$$\begin{array}{llll}
&\hskip2mm F  \approx f, \textmd{ if }  0< \vert f\vert \leq1; &
 F=0, \textmd{ if } f=0; &\hskip20mm  {\rm(3.7b)}
\\&&& \\&\hskip2mm
F\to -\ln (\ln 2+1),  \textmd{ as } f \to -\infty;   &    F \approx  \ln f \textmd{ as } f \to +\infty. & \hskip20mm  {\rm(3.7c)}
\end{array} $$
Conditions (3.7b) assure that the regularization is close to the original function in the domain of validity of KP , whereas    conditions (3.7c)  remove   negative infinity  and  reduce  large amplitudes to make them more physically plausible.

\begin{figure}[htp]
  \centerline{\includegraphics[scale=.25]{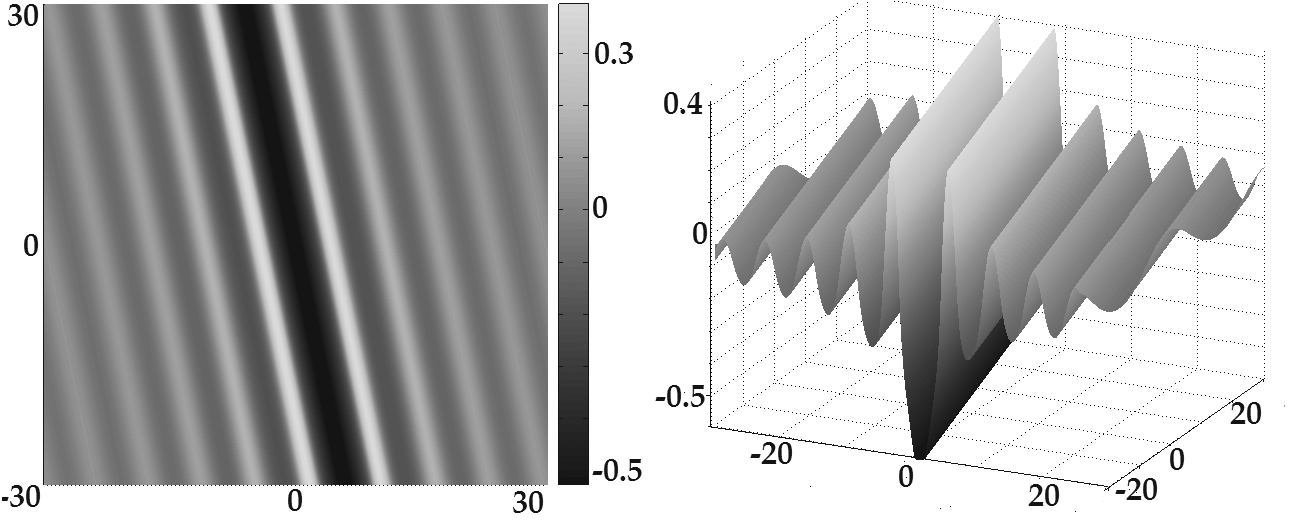}}% Images in 100% size
  \caption{Graph of regularization  (3.7) of (3.5) with $\alpha\hskip-0.5mm=\hskip-0.5mm1,  \;  \lambda\hskip-0.5mm=\hskip-0.5mm0.5,\;\mu\hskip-0.5mm=\hskip-0.5mm-0.1,\;
\gamma\hskip-0.5mm=\hskip-0.5mm0,\rho\hskip-0.5mm=\hskip-0.5mm0,  \; \chi\hskip-0.5mm=\hskip-0.5mm0,\; t\hskip-0.5mm=\hskip-0.5mm 0.  $ }
\label{fig:1}
\end{figure}

\begin{figure}[htp]
  \centerline{\includegraphics[scale=.25]{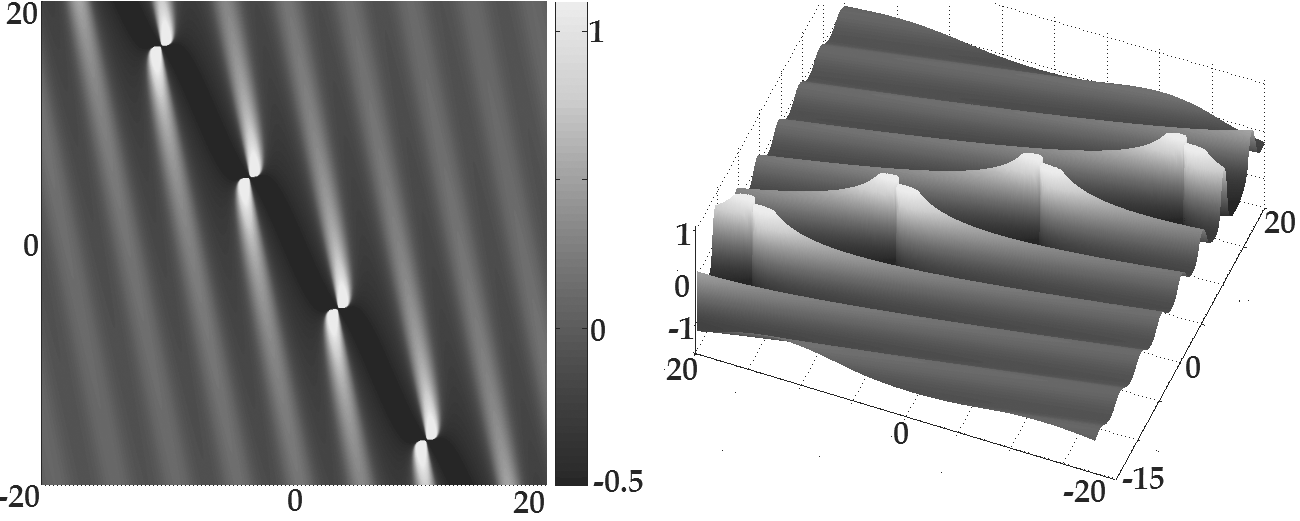}}% Images in 100% size
  \caption{Graph of regularization  (3.7) of (3.5) with $\alpha\hskip-0.5mm=\hskip-0.5mm1,  \,  \lambda\hskip-0.5mm=\hskip-0.5mm0.65,\,\mu\hskip-0.5mm=
  \hskip-0.5mm-0.1,\,\gamma\hskip-0.5mm=\hskip-0.5mm0,\,
  \rho\hskip-0.5mm=\hskip-0.5mm0,\,
 \chi\hskip-0.5mm=\hskip-0.5mm.105\pi,\, t\hskip-0.5mm=\hskip-0.5mm0. $ }
\label{fig:3}
\end{figure}

Regularization (3.7) of (3.5) with $\chi\hskip-1mm =\hskip-1mm0$ is shown in
Figure  \ref{fig:1}, while
regularization (3.7) of (3.5) with   $\chi\neq 0$ is  shown in   Figure  \ref{fig:3}.

\begin{figure}
  \centerline{\includegraphics[scale=.4]{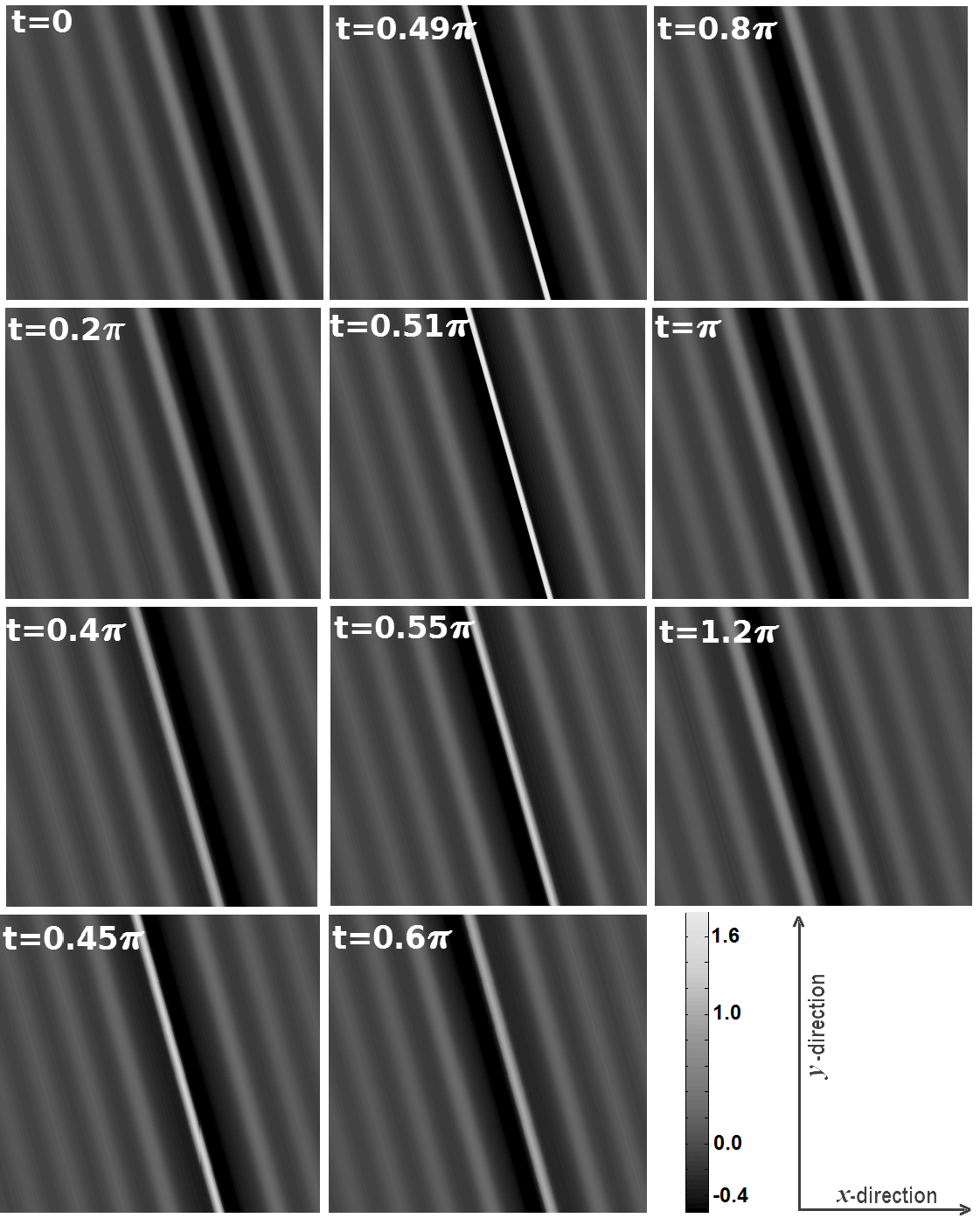}}% Images in 100% size
  \caption{Graph of regularization  (3.7) of (3.5) with $\alpha\hskip-0.5mm=\hskip-0.5mm1,  \;  \lambda\hskip-0.5mm=\hskip-0.5mm0.5,\;\mu\hskip-0.5mm=\hskip-0.5mm-0.1,\;
\gamma\hskip-0.5mm=\hskip-0.5mm0,\rho\hskip-0.5mm=\hskip-0.5mm0, \; \chi\hskip-0.5mm=\hskip-0.5mm0, -30<x<15 , -30<y<30, $ for the values of $t$ shown. }
\label{fig:harmbrea}
\end{figure}

We may determine the velocity $\pmb v=(v_x, v_y)$ of motion of the profile given by   (3.5)   by setting $\displaystyle\frac{\textmd{d}}{\textmd{d}t}\Big[  \lambda  x-2\lambda \mu y+4\lambda (\lambda ^2-3\alpha^2\mu ^2)t\Big]=\lambda v_ x-2\lambda \mu v_y+4\lambda (\lambda ^2-3\alpha^2\mu ^2)~=~0 $ and $\displaystyle\frac{\textmd{d}}{\textmd{d}t}\bigg[   \cos (\alpha\chi) x+2\Big[\displaystyle\frac{\lambda\sin (\alpha\chi)}{\alpha}-\mu\cos (\alpha\chi) \Big] y\enskip+  12\Big[\lambda^2\cos(\alpha\chi)-\alpha^2\mu^2\cos (\alpha\chi) +2\alpha \lambda\mu\sin (\alpha\chi) \Big]t\bigg] $ $=\cos (\alpha\chi) v_x+2\Big[\displaystyle\frac{\lambda\sin (\alpha\chi)}{\alpha}-\mu\cos (\alpha\chi) \Big] v_y\enskip+  12\Big[\lambda^2\cos(\alpha\chi)-\alpha^2\mu^2\cos (\alpha\chi) +2\alpha \lambda\mu\sin (\alpha\chi)=0, $ which give us    a  system of equations
$$\begin{array}{l}
   v_x  +2\Big[\displaystyle\frac{\lambda\tan(\alpha\chi)}{\alpha}-\mu  \Big]v_y+ 12\Big[\lambda^2 -\alpha^2\mu^2  +2\alpha \lambda\mu\tan (\alpha\chi)\Big]=0 \\\\
  v_x-2 \mu v_y+4 (\lambda^2-3\alpha^2\mu^2)=0.
\end{array}\eqno{\rm (3.8a)}$$
If $\lambda\tan  \alpha \chi\neq 0 $  the determinant of system (3.8a) is nonzero  and the  system has
  a unique solution
  $$
v_x=-12\alpha^2\mu^2-4\lambda^2 -\displaystyle\frac{8\alpha  \lambda \mu}{\tan \alpha \chi} ,\quad   v_y=-12\alpha^2\mu-\displaystyle\frac{4\alpha  \lambda }{\tan \alpha \chi} .
\eqno{\rm (3.8b)}
$$
Far away from the singular curve  the oscillatory portion of the waves moves in the direction $\left(\displaystyle\frac{1}{\sqrt{1+4\mu^2}} , \;\; \frac{ -2\mu }{\sqrt{1+4\mu^2}}\right) $ with the speed of  $\displaystyle\frac{v_x-2\mu v_y}{\sqrt{1+4\mu^2}}=\dfrac{ 12\alpha^2\mu^2-4\lambda^2}{\sqrt{1+4\mu^2}}$ which is independent of $\chi$. The singular curve itself stays within distance $\displaystyle \frac{1}{2  \big\vert \lambda \cos(\alpha \chi) \big\vert\; }$ from the straight line  $\displaystyle \frac{\rho}{\cos( \alpha\chi)} +x +2\Big[\displaystyle\frac{\lambda\tan (\alpha\chi)}{\alpha}-\mu \Big]y +  12\Big[\lambda^2 -\alpha^2\mu^2 +2\alpha \lambda\mu\tan (\alpha\chi)\Big]t=0$  which moves in the direction  of $\left(\displaystyle\frac{1}{\sqrt{1+4\big(\mu-\frac{\lambda \tan \alpha \chi}{\alpha}\big)^2}} ,  \frac{  2\left(\frac{\lambda \tan \alpha \chi}{\alpha}-\mu\right)}{\sqrt{1+4\big(\mu-\frac{\lambda \tan \alpha \chi}{\alpha}\big)^2}}\right) $ with the speed of  $\displaystyle -\frac{12\Big[\lambda^2 -\alpha^2\mu^2 +2\alpha \lambda\mu\tan (\alpha\chi)\Big]}{\sqrt{1+4\big(\mu-\frac{\lambda \tan \alpha \chi}{\alpha}\big)^2}}$ bounded for all values of $\chi$.
For $\chi=0$     profile (3.5) moves in the direction of $\left(\dfrac{1}{\sqrt{1+4\mu^2}},\dfrac{-2\mu}{\sqrt{1+4\mu^2}}\right)$ with the speed of $\displaystyle \dfrac{ 12(\alpha^2\mu^2- \lambda^2)}{\sqrt{1+4\mu^2}}$ while at the same time oscillating along the same direction.     The time evolution of such a  wave is shown in Figure \ref{fig:harmbrea}, it is similar for $\alpha=1$ and $\alpha=i.$

We
cannot attach a meaningful  definition  of velocity to the motion of nonlinear superposition of two  or more solutions (3.5). The velocity $\pmb v=(v_x, v_y)$ is not to be confused with the velocity of the flow given by the vector $\pmb u^\prime,$ the former is the velocity of motion of the profile while the latter is the velocity of motion of the fluid. The velocity field inside the fluid is  is always related to the pressure and at the surface the
pressure is constant (hydrostatic), while inside it has a complicated
behavior,   \cite{Constantin10}.

\begin{figure}
  \centerline{\includegraphics[scale=.25]{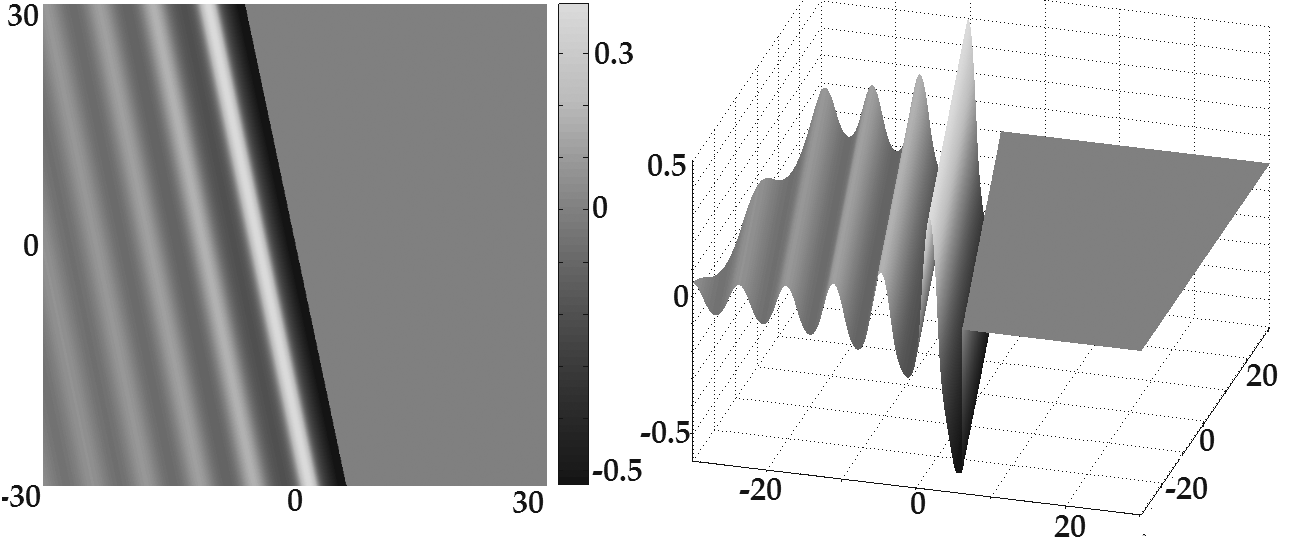}}% Images in 100% size
  \caption{Graph of regularization  (3.7) of  a simply connected component of (3.5) given by
   (3.9) with  $\alpha\hskip-0.5mm=\hskip-0.5mm1,  \;  \lambda\hskip-0.5mm=\hskip-0.5mm0.5,\;\mu\hskip-0.5mm=\hskip-0.5mm-0.1,\;
\gamma\hskip-0.5mm=\hskip-0.5mm0,\rho\hskip-0.5mm=\hskip-0.5mm0,\; \chi\hskip-0.5mm=\hskip-0.5mm0,\; t\hskip-0.5mm=\hskip-0.5mm 0.$  The plot is discontinuous  along the straight line   $2\lambda\Upsilon   - {\sin 2\Gamma }=0 $, however Matlab replaced the discontinuity   with a vertical strip and the author decided to leave it that way.}
\label{fig:2}
\end{figure}

\begin{figure}
  \centerline{\includegraphics[scale=.33]{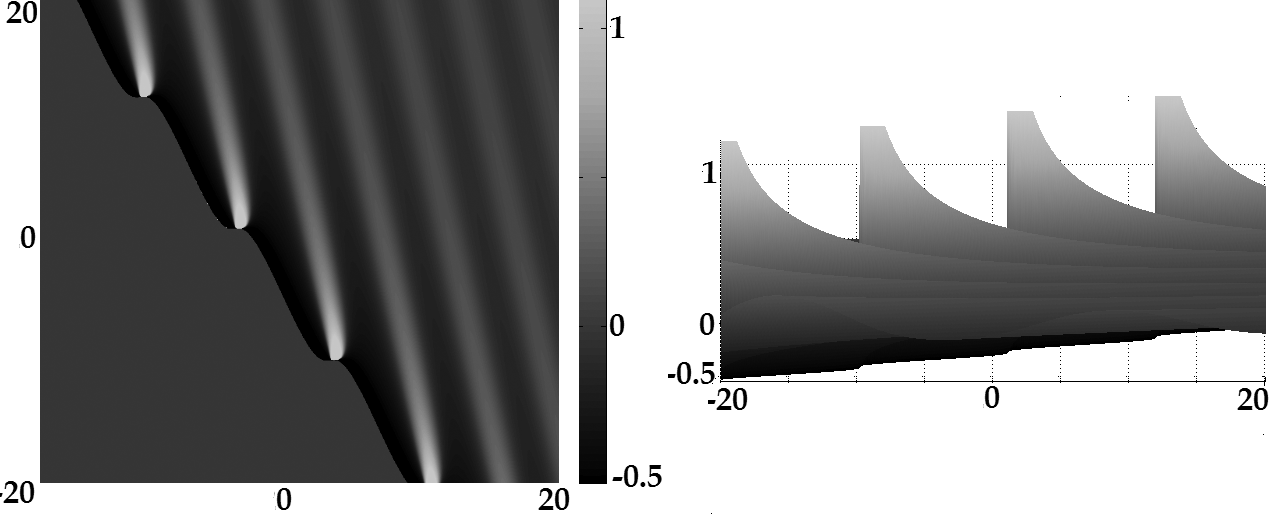}}% Images in 100% size
  \caption{Graph of regularization  (3.7) of  a simply connected component of (3.5) given by
   (3.9) with  $\alpha\hskip-0.5mm=\hskip-0.5mm1,  \;  \lambda\hskip-0.5mm=\hskip-0.5mm6.5,\;\mu\hskip-0.5mm=\hskip-0.5mm-0.1,\;
\gamma\hskip-0.5mm=\hskip-0.5mm0,\rho\hskip-0.5mm=\hskip-0.5mm0,\; \chi\hskip-0.5mm=\hskip-0.5mm0.105\pi ,\; t\hskip-0.5mm=\hskip-0.5mm 0.7 .$  The plot is discontinuous  along the curve  $2\lambda\Upsilon   - {\sin 2\Gamma }=0 $.}
\label{fig:figure3half}
\end{figure}

Each such solution consists of two   components separated by a singular line.  We may retain the component on one side of the singular line and set the function  to zero on the other side by taking
$$
\widetilde{f}(x,y,t)= \left\{
    \begin{array}{l}
      f(x,y,t)\textmd{ given by (3.5a), if } 2\lambda\Upsilon   - {\sin 2\Gamma } >0, \\[2pt]\\
    0, \textmd{ if }2\lambda\Upsilon   - {\sin 2\Gamma } <0.
  \end{array} \right. \eqno{\rm (3.9a)}
$$
or
$$
\widetilde{f}(x,y,t)= \left\{
    \begin{array}{l}
      f(x,y,t)\textmd{ given by (3.5), if } 2\lambda\Upsilon   - {\sin 2\Gamma } <0, \\[2pt]\\
    0, \textmd{ if }2\lambda\Upsilon   - {\sin 2\Gamma } >0.
  \end{array} \right. \eqno{\rm (3.9b)}
$$
Sample graphs of (3.9) are shown in Figure  \ref{fig:2} for $\chi=0$ and in Figure  \ref{fig:figure3half} for $\chi \neq0$.

\begin{figure}
  \centerline{\includegraphics[scale=.21]{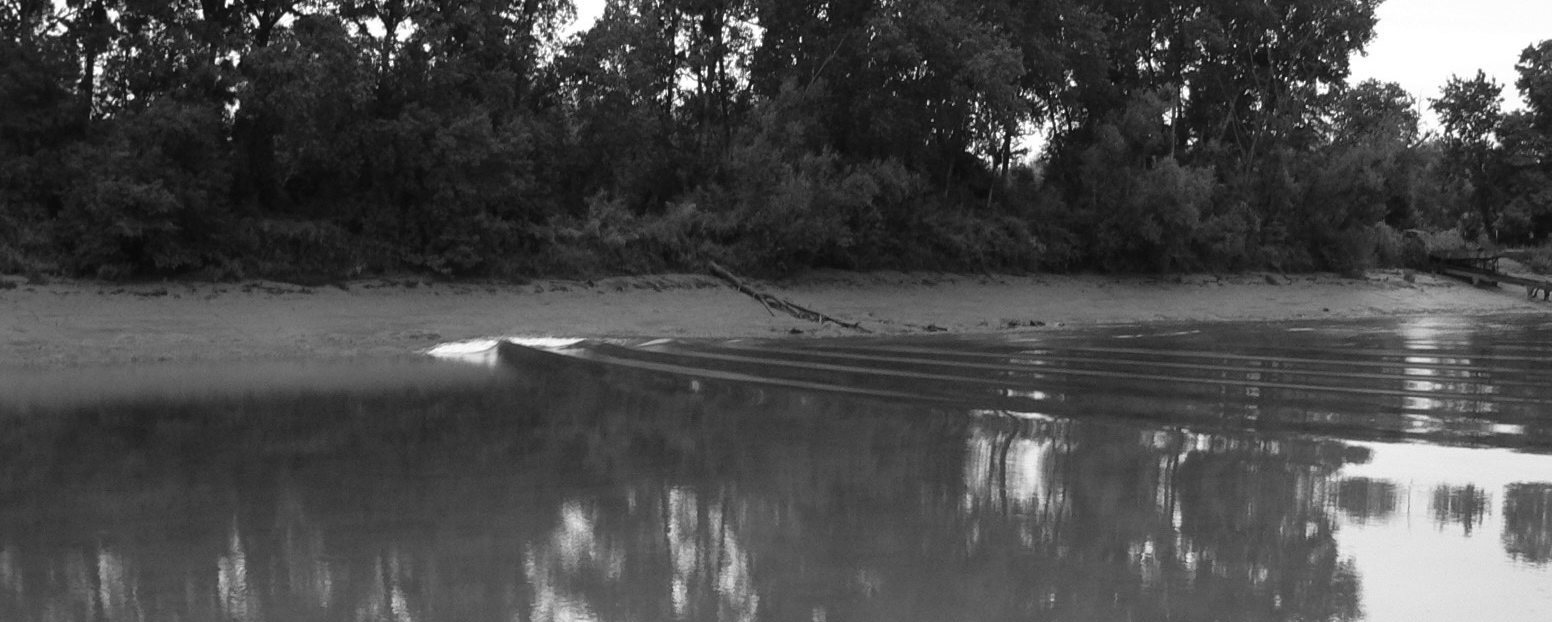}}% Images in 100% size
  \caption{ Garonne river tidal bore at Arcins on  July 6, 2008 around 07:10 (views from right bank), described at  \cite{crsl}.   Courtesy of Hubert Chanson.   }
\label{fig:undularbore}
\end{figure}

For $\chi=0$ the obtained function is singular along the line $  2\lambda\Upsilon   - {\sin 2\Gamma } =0$ and thus could not be expected to model any physical wave in a
neighborhood $\Big\vert 2\lambda\Upsilon   - {\sin 2\Gamma } <0\Big\vert < \varepsilon,  \; \varepsilon$ is some constant, of the singular line.  But away from the singular line in the domain $\Big\vert 2\lambda\Upsilon   - {\sin 2\Gamma } <0\Big\vert > \varepsilon $  function   $\widetilde{f}(x,y,t)$ provides a quite good description of {\it undular bores } or waves with the advancing strongly-pronounced front followed by a train of well-defined free-surface undulations. Near the singular line  $\widetilde{f}(x,y,t) $  has a large negative trough followed  by a large positive crest as shown in Figure  \ref{fig:2}.  Formula (2.43) for the physical velocity  $   u^\prime_{x^\prime} =\epsilon\sqrt{gh}\;\eta_0+o(\epsilon ) $ shows that while  water at the crest of the wave  continues to move forward, water in the trough moves backwards  leading to  wave  overturning which cannot be described within the framework of KP. Numerous   discussions and pictures of undular bores may be found on the Internet, e. g.  \cite{ w3},  one of which is  reproduced
  in Figure  \ref{fig:undularbore}.  Some of the best places to observe undular bores are the French river Seine where the bores form  on spring tides and reach  some 50 km inland,   the Fu-chun River in China, the Amazon River in Brasil, the Severn in England and the Petitcodiac River in New Brunswick, Canada.

\begin{figure}
  \centerline{\includegraphics[scale=.28]{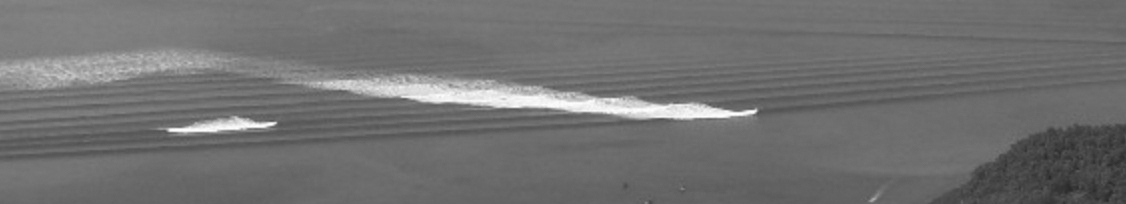}}% Images in 100% size
  \caption{ The tsunami of December 26, 2004 approaching Koh Jum island, off the coast of Thailand, after it passed around the Phi
Phi islands.   Copyright Anders Grawin, 2006. Reproduced from http://www.kohjumonline.com/anders.html with permission. }
\label{fig:anders}
\end{figure}

An undular bore-like structure was also observed in December 26, 2004 tsunami. As the tsunami passed around the Phi
Phi Islands and moved towards Koh Jum island off the cost of Thailand  it took undular bore-like form and was photographed by Anders Grawin. The pictures are reproduced in \cite{ Constantin08}    and \cite{ Anders04}  with  one of them shown in Figure  \ref{fig:anders}.

\begin{figure}
\center{ {\includegraphics[scale=.5]{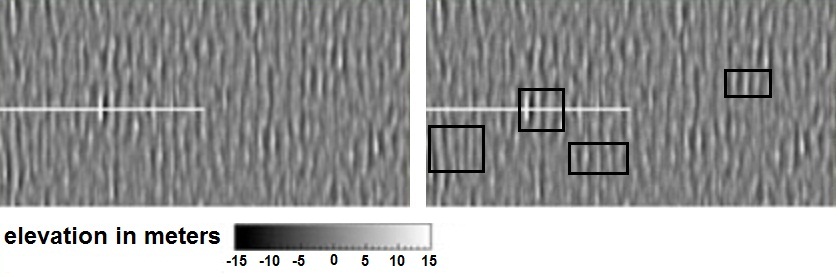}}}% Images in 100% size
  \caption{Image  captured by ESA, courtesy of ESA and Deutsches Zentrum fur Luft- und Raumfahrt (DLR).  On the left is the original image, on the right is the same image with some waves of the type  shown in Figures  \ref{fig:1} and  \ref{fig:2} enclosed in black rectangles.  The middle of the horizontal  white line marks the wave with the highest crest,  it is the leading crest of a wave of the type shown in Figure  \ref{fig:2}. \hskip11cm }
\label{fig:w17a0}\end{figure}

\begin{figure}
\center{ {\includegraphics[scale=.4]{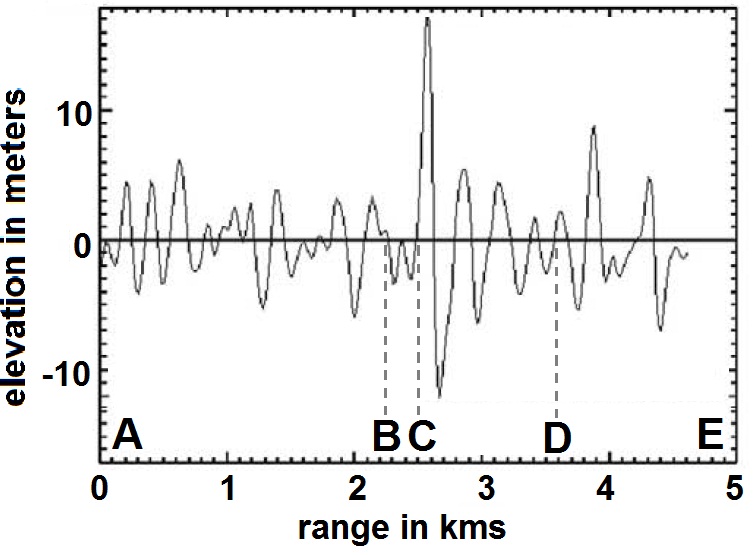}}}% Images in 100% size
  \caption{ The vertical transect of the ocean surface in Figure \ref{fig:w17a0} along the straight white line.  The wave amplitude in   region CD is as predicted by (3.9) and Figure  \ref{fig:2},  region BC corresponds to the region where (3.9)  exhibits  negative singularity. In regions AB and DE the undular bore-like behavior is masked  by other waves.   \hskip11cm }
\label{fig:w18}\end{figure}

Several undular bore-like waves are shown in Figure \ref{fig:w17a0} in the  image   captured by ESA, with some analysis of one of these waves shown in Figure \ref{fig:w18}.

%The picture was emailed to me May 1, 2012 at 5:43 am by Hubert Chanson

%The permission was granted  by Roger Brooke, see email exchange in the %email from Roger Brooke received April 28, 2012 at 10:13 pm.

%Permission was granted by Susanne Lehner on behalf of    Deutsches Zentrum fur Luft- und Raumfahrt
%  in email received May 12 2012 at 2:46 am

%Permission was granted by Susanne Lehner on behalf of    Deutsches Zentrum fur Luft- und Raumfahrt
%  in email received May 12 2012 at 2:46 am

The very existence  of the singular line indicates that  most of the power of an undular bore is concentrated in the trough at the leading edge rather than at the  bore's crests. The presence  of large forces at the leading edge of   undular bores  is described in  \cite{Simpson04, Mouaza10, Wolanski04}.
 The account of an undular bore  in  \cite{Wolanski04}  as "  unsteady motion ... sufficiently energetic to topple moorings that had survived much higher, quasi-steady currents of 1.8 m/s " also supports the claim  that the real power of an undular bore is in its leading trough rather than relatively modest crests even though the leading trough may not be well-pronounced.   As follows from (3.5) the leading crest may change its shape.

 Often the undular bore-like waves are misinterpreted as   trains of solitons with diminishing amplitudes. If that were true, the solitons would be dispersing as the smaller ones travel slower than the taller ones.  Yet the undular bore-like structures do not disperse and preserve their shapes for quite a while behaving more like solutions (3.5) than trains of solitons. We  refer to solutions (3.5) as "harmonic breathers".

 Although most undular bores-like waves have the shape shown in Figure \ref{fig:2} with the strongest crest-trough pair leading the wave, there might have been observations of waves with the shape resembling Figure   \ref{fig:1} with the strongest crest-trough pair being in the middle rather than in the lead of the wave.   According to  \cite{Dudley},   the Aleutian tsunami of
1946 produced  multi-crest waves  with
the third or fourth crests being the highest
and most violent, and a bore at the Waimea  Riverr on Kaua'i is reported to have the sixth crest as
the highest and most destructive.

\begin{figure}
  \centerline{\includegraphics[scale=.43]{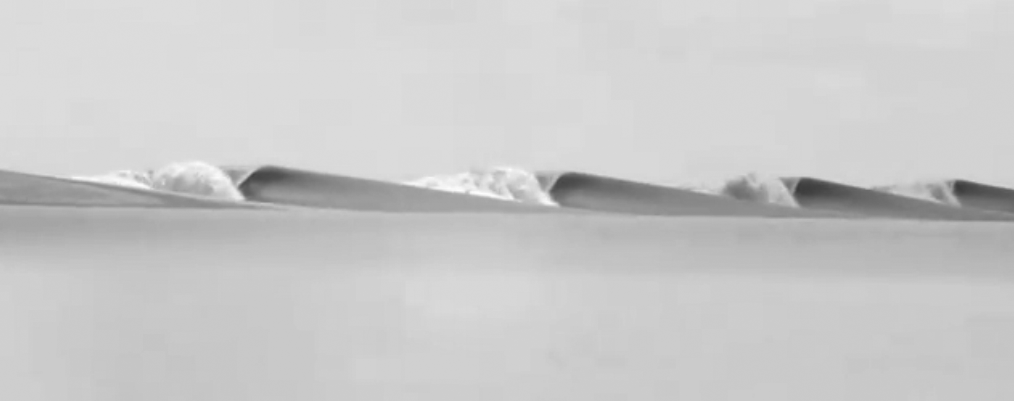}}% Images in 100% size
  \caption{Time frame $t=1:21$ from the video at \cite{w5}. The shape of the wave is in agreement with Figure \ref{fig:figure3half} except for   wave overturning exhibited by the physical wave shown but not by Figure \ref{fig:figure3half}.  }
\label{fig:7ghosts1}
\end{figure}

\begin{figure}
  \centerline{\includegraphics[scale=.6]{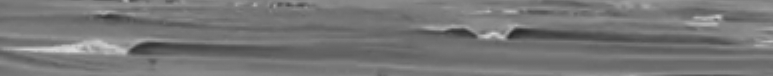}}% Images in 100% size
   \caption{Portion of time frame $t=2:11$ from the video at \cite{w5}.  The shape of the wave is in agreement with Figure \ref{fig:3}  with $\chi$ close to but not equal to zero except for   wave overturning exhibited by the physical wave shown but not by Figure \ref{fig:3}. }
\label{fig:7ghosts0}
\end{figure}

For $\chi\neq0$  equation   $  2\lambda\Upsilon   - {\sin 2\Gamma } =0$ describes a curve rather than a straight line, along the curve  function (3.5) is singular and   could not be expected to model any physical wave in a
neighborhood $\Big\vert 2\lambda\Upsilon   - {\sin 2\Gamma } <0\Big\vert < \varepsilon,  \; \varepsilon$ is some constant, of the singular curve.  However,  away from the singular curve  the function describes the waves observed in the river Seven Ghosts  in Sumatra, Indonesia and shown in Figures  \ref{fig:7ghosts1}, \ref{fig:7ghosts0} taken from the video  \cite{w5}.

%Permission was granted by Scott McClimont Video Content Manager of
% Rip Curl in email received April 11, 2012 at 8:10 am
%Permission was granted by Scott McClimont Video Content Manager of
% Rip Curl in email received April 11, 2012 at 8:10 am

  Formula (2.43) for the physical velocity  $   u^\prime_{x^\prime} =\epsilon\sqrt{gh}\;\eta_0+o(\epsilon ) $ shows that while  water in the trough moves backwards, at the points behind the singular curve where the water level is relatively high water continues  to move forward,   leading to  wave  overturning but only at some points along the singular curve in agreement with  the video  taken at the river  Seven Ghosts.

Another class of explicit solutions of KP   may be obtained by  simple  formal substitutions
 $$  \begin{array}{lll}
&  \lambda  \to i\lambda  , \;\; \;\;\chi  \to i\chi    , \;\; \;\;\gamma  \to i\gamma \;\;\;\textmd{ into (3.5)}\\ &\\
&   \lambda_n \to i\lambda_n , \; \chi_n \to i\chi_n   , \; \gamma_n \to i\gamma_n  \textmd{ into (3.6)}
\end{array}\eqno{\rm (3.10)} $$
Substitutions (3.10) give us
  $$
\begin{array}{lll}
&f(x,y,t)=2\displaystyle\frac{\partial^2}{\partial x^2}\ln  \Big[2\lambda\Upsilon   - {\sinh 2\Gamma } \Big],& {\rm(3.11a)}\\&& \\
&\Upsilon =\rho +x\cosh (\alpha\chi) -2\Big[\displaystyle\frac{\lambda\sinh (\alpha\chi)}{\alpha}+\mu\cosh (\alpha\chi) \Big]y\enskip-&\\&&\\
&\hskip2.1cm 12\Big[\lambda^2\cosh(\alpha\chi)+\alpha^2\mu^2\cosh (\alpha\chi) +2\alpha \lambda\mu\sinh (\alpha\chi)\Big]t,\hskip7mm
 &  {\rm(3.11b)} \\&&\\
&\Gamma=  \gamma+\lambda x-2\lambda\mu y-4\lambda(\lambda^2+3\alpha^2\mu^2)t,
& {\rm(3.11c)}
\end{array}
$$
where  $\lambda, \mu, \chi, \gamma, \rho  $ are some  constants. If the constants   $\lambda, \mu, \chi, \gamma, \rho  $ are real so is the solution (3.5).  Nonlinear superposition of such solutions is of the form
$$
u(x,y,t)=2\frac{\partial^2}{\partial x^2}\ln  \det{\pmb  K}
\eqno{{\rm(3.12a)}}
$$
where ${\pmb  K}$ is an $N\times N$ matrix with the entries
$$
{\pmb  K}=\left(
\begin{array}{ccccc}
K_{11}&K_{12} &\dots& K_{1N} \\
K_{21} &K_{22}&\dots& K_{2N} \\
\vdots&\vdots&\vdots&\vdots \\
K_{N1} &K_{N2} &\dots &K_{NN}
\end{array}\right)   \eqno{\rm (3.12b)}
$$
$$
\begin{array}{lll}
&K_{nn}= \Upsilon_n  -\displaystyle\frac{\sinh 2\Gamma_n}{2\lambda_n}\enspace,
  & {\rm (3.12c)}                 \\ &&\\
&K_{nk}=   \left[-
\displaystyle\frac{(\lambda_n-\lambda_k)\sinh(\Gamma_n-\Gamma_k)}{\alpha^2(\mu_n-\mu_k)^2-
(\lambda_n-\lambda_k)^2}+ \displaystyle\frac{(\lambda_n+\lambda_k)\sinh(\Gamma_n+\Gamma_k)}{\alpha^2 (\mu_n-\mu_k)^2-(\lambda_n+\lambda_k)^2}\right]
+ & \\&& \\
  &\hskip1mm  \alpha     \left[  \displaystyle\frac{(\mu_n-\mu_k)\cosh(\Gamma_n+\Gamma_k)}{\alpha^2(\mu_n-\mu_k)^2-(\lambda_n+\lambda_k)^2}
   -
\displaystyle\frac{(\mu_n-\mu_k)\cosh(\Gamma_n-\Gamma_k)}{\alpha^2 (\mu_n-\mu_k)^2-(\lambda_n-\lambda_k)^2}\right], \enspace n\ne k
  &  {\rm (3.12d)}  \\&& \\
&\Upsilon_n=\rho_n+x\cosh (\alpha\chi_n) -2\Big[\displaystyle\frac{\lambda_n\sinh (\alpha\chi_n)}{\alpha}+\mu_n\cosh (\alpha\chi_n) \Big]y\enskip-&\\
&\hskip0.9cm 12\Big[\lambda_n^2\cosh(\alpha\chi_n)+\alpha^2\mu_n^2\cosh (\alpha\chi_n) +2\alpha \lambda_n\mu_n\sinh (\alpha\chi_n)\Big]t,
 &{\rm(3.12e)} \\ \\
&\Gamma_n=  \gamma_n+\lambda_n x-2\lambda_n\mu_n y-4\lambda_n(\lambda_n^2+3\alpha^2\mu_n^2)t,
 &{\rm(3.12f)}
\end{array}
$$
where $\lambda_n\textmd{'s,}$$\mu_n\textmd{'s,}$$\chi_n\textmd{'s, }$$\gamma_n\textmd{'s, }$$\rho_n\textmd{'s, }$ are   constants.  Change of constants $\rho_n\to -\rho_n, \, \chi_n\to  \displaystyle \frac{\chi_n}{\alpha}+i\pi$ allows us to change the sign in front of $\Upsilon_n$ while the change $\gamma_n\to \gamma_n\pm \displaystyle i\frac {\pi}{2}$ for all $n=1,2,3\dots,N$ allows us to change the signs in front of all   $\displaystyle\frac{(\lambda_n+\lambda_k)\sinh(\Gamma_n+\Gamma_k)}{\alpha^2 (\mu_n-\mu_k)^2-(\lambda_n+\lambda_k)^2}, \; \displaystyle\frac{(\mu_n-\mu_k)\cosh(\Gamma_n+\Gamma_k)}{\alpha^2(\mu_n-\mu_k)^2-(\lambda_n+\lambda_k)^2}.$

 \begin{figure}
  \centerline{\includegraphics[scale=.25]{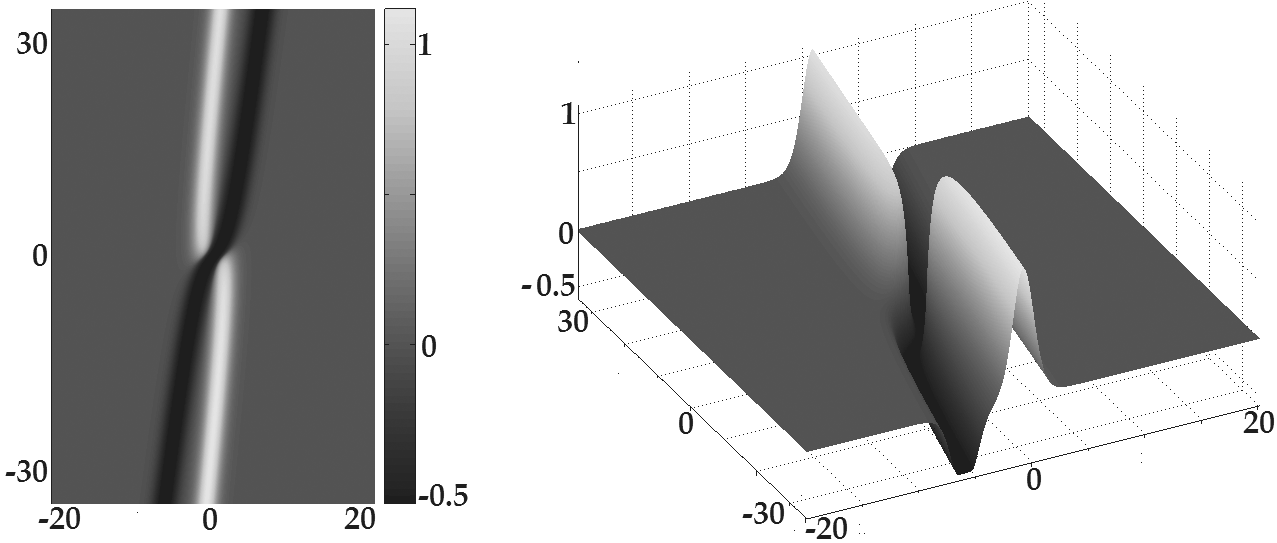}}% Images in 100% size
  \caption{Graph of regularization  (3.7) of (3.11) with $\alpha\hskip-0.5mm=\hskip-0.5mm1,  \,  \lambda\hskip-0.5mm=\hskip-0.5mm1,\,\mu\hskip-0.5mm=\hskip-0.5mm0.05,\,
  \gamma\hskip-0.5mm=\hskip-0.5mm0,\,\rho\hskip-0.5mm=\hskip-0.5mm0,
\, \chi\hskip-0.5mm=\hskip-0.5mm0.6,\, t\hskip-0.5mm=\hskip-0.5mm0. $ }
\label{fig:4plus}\end{figure}

\begin{figure}
  \centerline{\includegraphics[scale=.25]{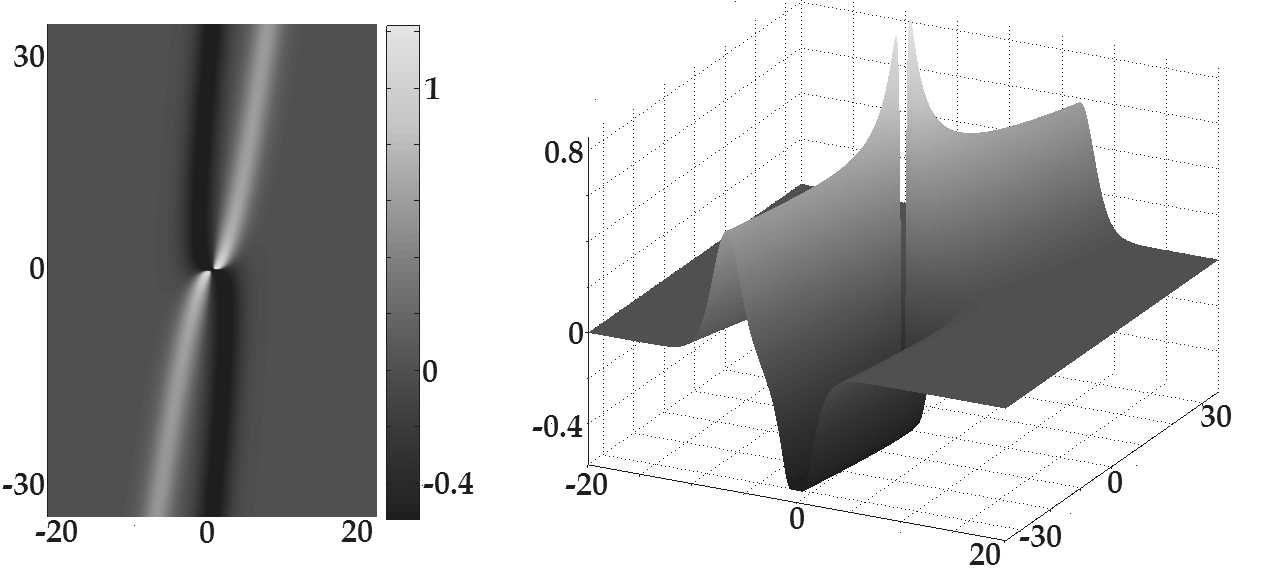}}% Images in 100% size
  \caption{Graph of regularization  (3.7) of (3.11) with $\alpha\hskip-0.5mm=\hskip-0.5mm i,  \,  \lambda\hskip-0.5mm=\hskip-0.5mm0.6,\,\mu\hskip-0.5mm=\hskip-0.5mm0.05,\,
\gamma\hskip-0.5mm=\hskip-0.5mm0,\,\rho\hskip-0.5mm=\hskip-0.5mm0,\, $  \protect\\
      $\chi\hskip-0.5mm=\hskip-0.5mm0.6,\, t\hskip-0.5mm=\hskip-0.5mm0.
  \hskip110mm$ }
\label{fig:4minus}\end{figure}

\begin{figure}
  \centerline{\includegraphics[scale=.40]{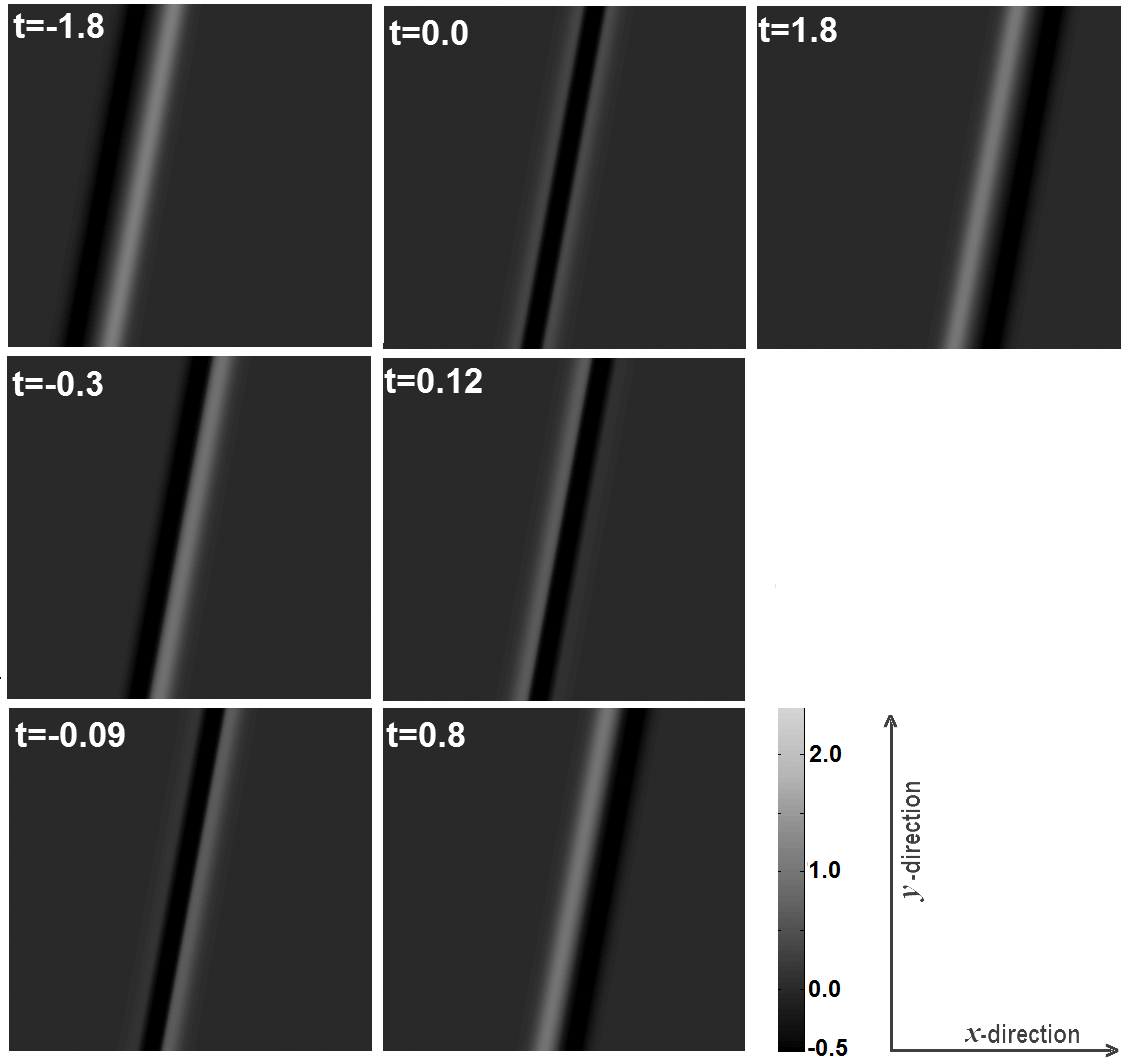}}% Images in 100% size
  \caption{Graph of regularization  (3.7) of (3.11) with $\alpha\hskip-0.5mm=\hskip-0.5mm1,  \;  \lambda\hskip-0.5mm=\hskip-0.5mm1,\;\mu\hskip-0.5mm=\hskip-0.5mm-0.05,\;
\gamma\hskip-0.5mm=\hskip-0.5mm0,\rho\hskip-0.5mm=\hskip-0.5mm0, \; \chi\hskip-0.5mm=\hskip-0.5mm0, -20<x<20 , -35<y<35, $ for the values of $t$ shown. }
\label{fig:figure4plus0}
\end{figure}

\begin{figure}
  \centerline{\includegraphics[scale=.40]{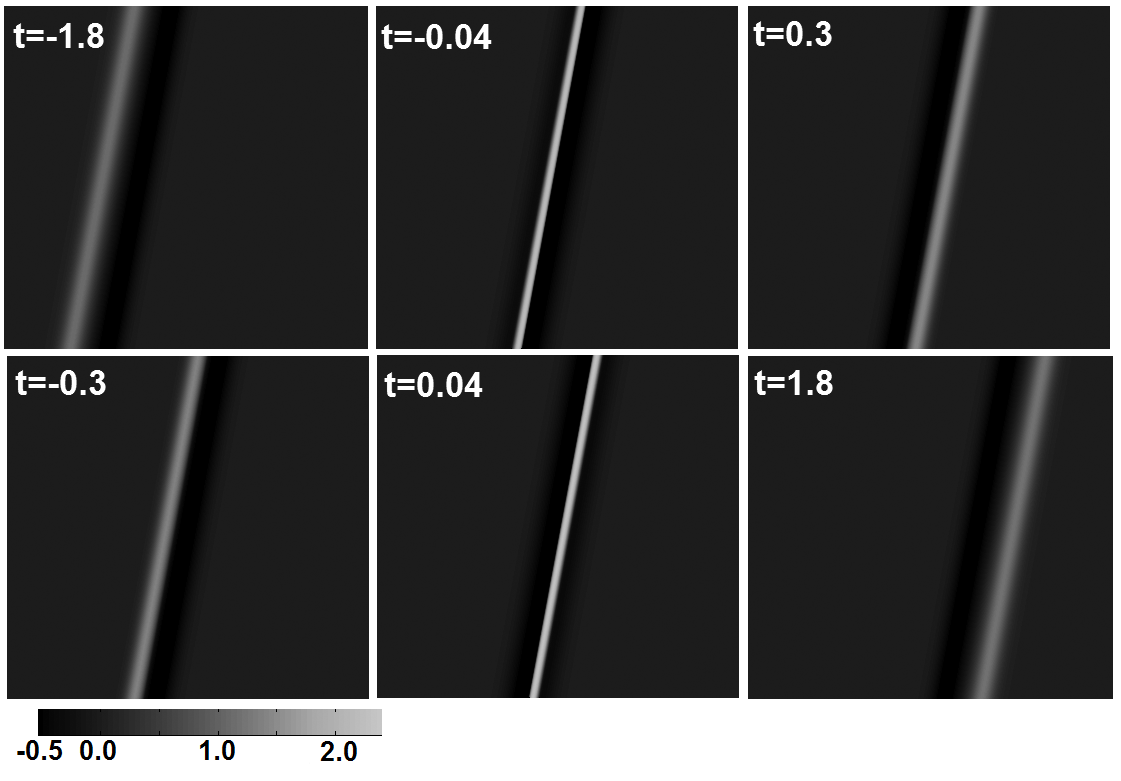}}% Images in 100% size
  \caption{Graph of regularization  (3.7) of (3.11) with $\alpha\hskip-0.5mm=\hskip-0.5mmi,  \;  \lambda\hskip-0.5mm=\hskip-0.5mm1,\;\mu\hskip-0.5mm=\hskip-0.5mm-0.05,\;
\gamma\hskip-0.5mm=\hskip-0.5mm0,\rho\hskip-0.5mm=\hskip-0.5mm0, \; \chi\hskip-0.5mm=\hskip-0.5mm0, -20<x<20 , -35<y<35, $ for the values of $t$ shown.\hskip7.0cm}
\label{fig:figure4minus0}
\end{figure}

Due to (2.43) the fluid moves to the right whenever $f>0,$ such regions are shown in Figures \ref{fig:4plus}, \ref{fig:4minus}    in light shades; the fluid moves  to the left whenever $f<0$, such regions are shown in Figures \ref{fig:4plus},  \ref{fig:4minus} in dark shades. When for $\chi\neq 0$ the fluid moving to the right collides with the fluid moving to the left, we obtain a "crossing" of light and dark lines. The velocity $\pmb v=(v_x, v_y)$ of the motion of the "crossing"   is the same as the velocity of the wave profile and its components may be uniquely determined from the system of equations
$$\begin{array}{ll}
&   v_x  -2\Big[\displaystyle\frac {\lambda   \tanh \alpha \chi }{\alpha} + \mu  \Big]v_y= 12\Big[ \lambda^2  +\alpha^2 \mu^2 +2  \lambda\mu \alpha  \tanh\alpha \chi \Big]  \\ &\\&
  v_x-2 \mu v_y= 4 (\lambda^2+3\alpha^2 \mu^2)
\end{array}\eqno{\rm (3.13a)}$$
obtained from (3.8a)  by means of substitutions (3.7). If the determinant of system (3.8a) is nonzero, i. e. $\lambda \tanh \alpha  \chi \neq 0$, the system has
a unique solution
$$
v_x=4\lambda^2-12\alpha^2\mu^2-\displaystyle\frac{8\lambda\mu \alpha}{\tanh \alpha \chi},\quad   v_y= -\displaystyle\frac{4\lambda\alpha }{\tanh \alpha \chi}-12\alpha^2\mu.
\eqno{\rm (3.13b)}
$$
The velocity component $\displaystyle\frac{v_x-2\mu v_y}{\sqrt{1+4\mu^2}}=\dfrac{ 12\alpha^2\mu^2+4\lambda^2}{\sqrt{1+4\mu^2}}$ is independent of $\chi$ while the component  $\displaystyle\frac{2\mu v_x+ v_y}{\sqrt{1+4\mu^2}}$ becomes infinite if $ \chi=0.$    For $\chi\neq0$   the motion of profile (3.11)  may be viewed as superposition of two motions: one is the {\it transverse motion }  in the direction of $\left(\dfrac{1}{\sqrt{1+4\mu^2}},-\dfrac{2\mu}{\sqrt{1+4\mu^2}}\right)$ with the speed of $\displaystyle \dfrac{ 12\alpha^2\mu^2+4\lambda^2}{\sqrt{1+4\mu^2}}$ and the other one is the {\it lateral motion } of the "crossing"  along the wave in the direction of  $\left(\dfrac{2\mu}{\sqrt{1+4\mu^2}}, \dfrac{1}{\sqrt{1+4\mu^2}}\right)$  with velocity $\displaystyle\frac{2\mu v_x+ v_y}{\sqrt{1+4\mu^2}}$ .
 For $\chi=0$     profile (3.11) moves in the direction of $\left(\dfrac{1}{\sqrt{1+4\mu^2}},-\dfrac{2\mu}{\sqrt{1+4\mu^2}}\right)$ with the speed of $\displaystyle \dfrac{ 12\alpha^2\mu^2+4\lambda^2}{\sqrt{1+4\mu^2}}$ while at the same time the crest grows,  jumps over the  singular line and then disappears, in this case the velocity of the  lateral component of the motion becomes infinite.    The time evolution of such  waves for $\alpha=1$ and $\alpha=i$ is shown in Figures \ref{fig:figure4plus0} and \ref{fig:figure4minus0}

We
cannot attach a meaningful  definition  of velocity to the motion of nonlinear superposition of two  or more solutions (3.11). The velocity $\pmb v=(v_x, v_y)$ is not to be confused with the velocity of the flow given by the vector $\pmb u^\prime,$ the former is the velocity of motion of the profile while the latter is the velocity of motion of the fluid.

\begin{figure}
  \centerline{\includegraphics[scale=.25]{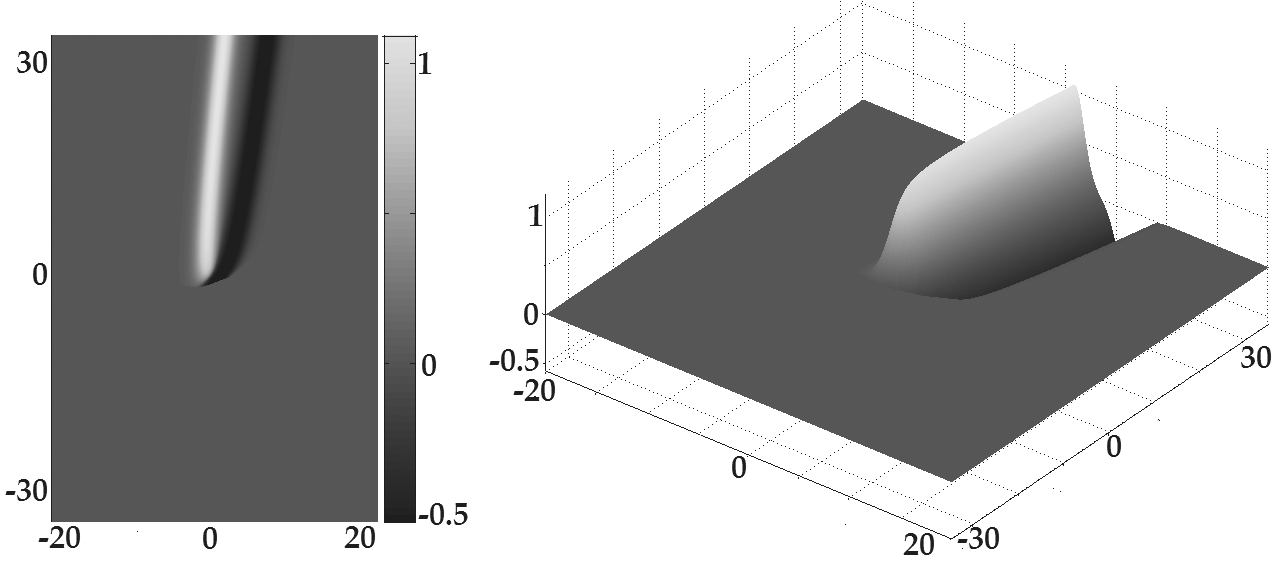}}% Images in 100% size
  \caption{Graph of regularization  (3.7) of the upper half of (3.11) with $\alpha\hskip-0.5mm=\hskip-0.5mm 1,  \,  \lambda\hskip-0.5mm=\hskip-0.5mm1,\,\mu\hskip-0.5mm=\hskip-0.5mm0.05,\,
 \gamma\hskip-0.5mm=\hskip-0.5mm0,\,\rho\hskip-0.5mm=\hskip-0.5mm0,\,  \chi\hskip-0.5mm=\hskip-0.5mm0.6,\, t\hskip-0.5mm=\hskip-0.5mm0.
  $  We deliberately do not specify how the upper half of the function shown in Figure 12 was cut out of the whole function; since it is not clear how to do it properly we used an ad-hoc way.}
\label{fig:5plus}\end{figure}

\begin{figure}
  \centerline{\includegraphics[scale=.25]{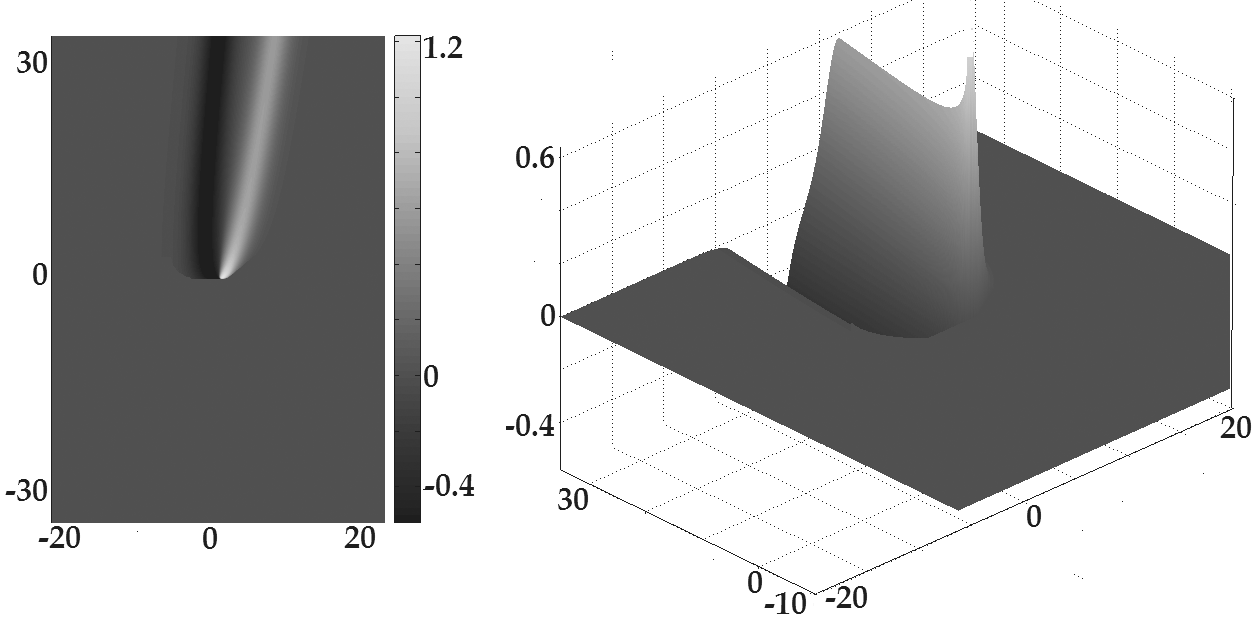}}% Images in 100% size
  \caption{Graph of regularization  (3.7) of the upper half of (3.11) with $\alpha\hskip-0.5mm=\hskip-0.5mm i,  \,  \lambda\hskip-0.5mm=\hskip-0.5mm0.6,\,\mu\hskip-0.5mm=\hskip-0.5mm0.05,\,
  \gamma\hskip-0.5mm=\hskip-0.5mm0,\,\rho\hskip-0.5mm=\hskip-0.5mm0,\, \chi\hskip-0.5mm=\hskip-0.5mm0.6,\, t\hskip-0.5mm=\hskip-0.5mm0.
 $We deliberately do not specify how the upper half of the function shown in Figure 13 was cut out of the whole function; since it is not clear how to do it properly we used an ad-hoc way.}
\label{fig:5minus}\end{figure}

\begin{figure}
 {\includegraphics[scale=.30]{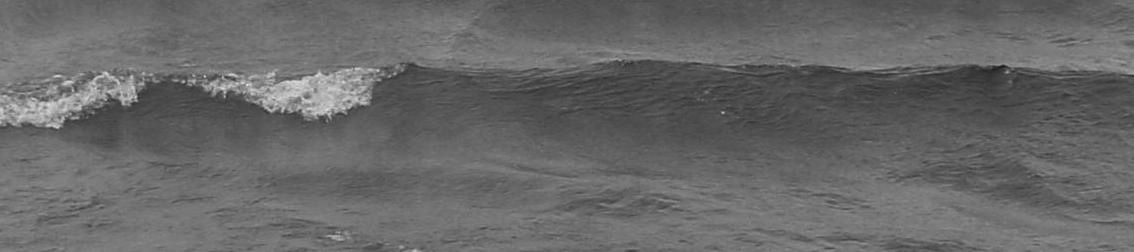}}% Images in 100% size
  \caption{  Physical wave  with the shape   predicted by Figures \ref{fig:5plus} - \ref{fig:5minus}.}
\label{fig:6}\end{figure}

\begin{figure}
\center{ {\includegraphics[scale=.5]{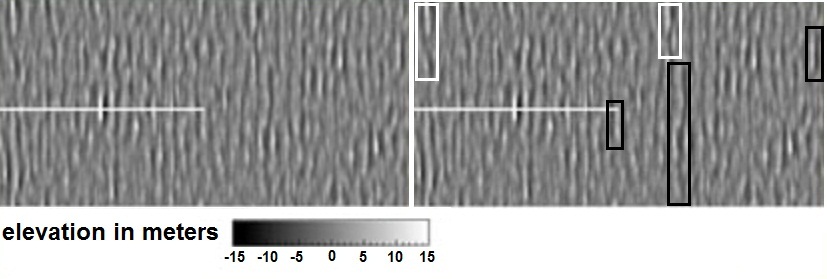}}}% Images in 100% size
  \caption{Image  captured by ESA, courtesy of ESA and Deutsches Zentrum fur Luft- und Raumfahrt (DLR).  On the left is the original image, on the right is the same image with some waves of the type  shown in Figures  \ref{fig:figure4plus0}, \ref{fig:5plus} enclosed in black rectangles and two waves  of the type shown in  Figure  \ref{fig:4plus} enclosed in white rectangles. In the latter   the "crossings" can be clearly seen.    \hskip11cm }
\label{fig:w17}\end{figure}

 Each of the Figures \ref{fig:4plus}, \ref{fig:4minus}  consists of two long crests   preceded or followed by long troughs, each crest-trough pair may be viewed as a separate wave looking somewhat like what is shown in Figures \ref{fig:5plus}, \ref{fig:5minus}. The physical phenomena described by such functions are waves that stretch along an almost straight half-line, with amplitude sharply dropping at the end point and slowly decaying to zero in the other direction, one such wave shown in Figure \ref{fig:6}.
 Much better illustrations may be found at numerous web sites, e.g. \cite{ w7, w8}. An image obtained by    ESA from a  satellite  and reproduced in Figure \ref{fig:w17} exhibits  several such waves, each is shown by a pair of white and black strips next to each other. One such wave is  described in  \cite{Graham2000a}     as " Only the very long
swell, of about 15 feet high and probably 600 to 1000 feet long. \dots   We were on the wing of the bridge, with a height of eye of 56 feet, and this wave broke over our heads. \dots  we were diving down off the face of the second of a set of three waves, so the ship just
kept falling into the trough, which just kept opening up under us."


% The picture was taken by the author

Although some of the literature  refer to such waves as
  "rogue waves", we   refrain from using the term for this type of waves as typically such waves appear in "predictable"  situations unlike real rogue waves that appear seemingly out of nowhere. While the sharp crest of the wave is clearly visible the trough may not be as apparent.

%Permission was granted by Susanne Lehner on behalf of    Deutsches Zentrum fur Luft- und Raumfahrt
%  in email received May 12 2012 at 2:46 am

\begin{figure}
  \centerline{\includegraphics[scale=.27]{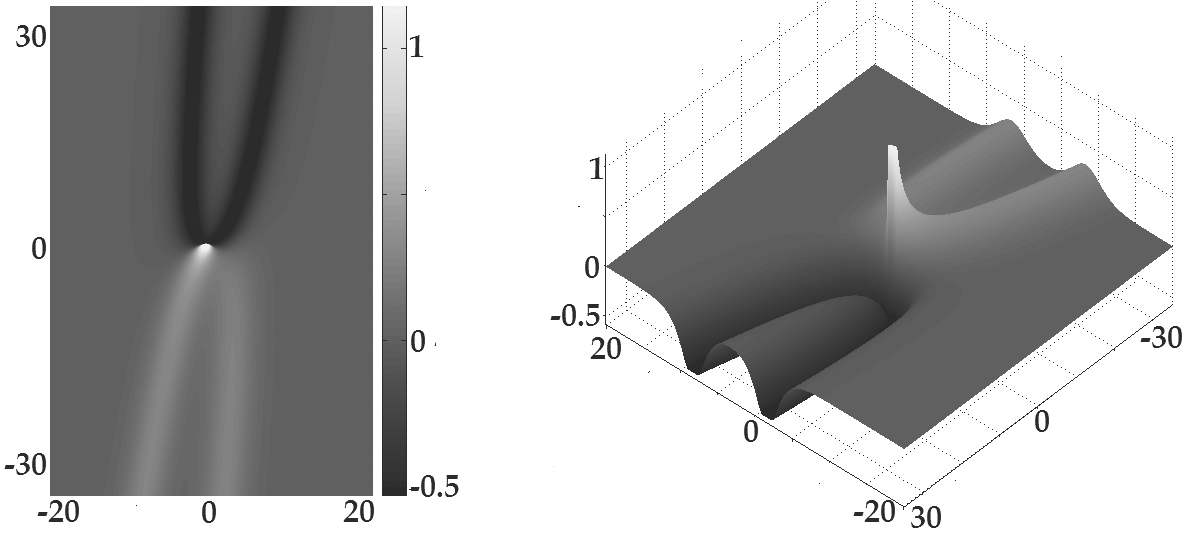}}% Images in 100% size
  \caption{Graph of regularization  (3.7) of  (3.15) with $\alpha\hskip-0.5mm=\hskip-0.5mm 1,  \,  \lambda\hskip-0.5mm=\hskip-0.5mm0.4,\,\mu\hskip-0.5mm=\hskip-0.5mm0.05,\,
\gamma\hskip-0.5mm=\hskip-0.5mm0, \,\rho\hskip-0.5mm=\hskip-0.5mm0, , \chi\hskip-0.5mm=\hskip-0.5mm0.6,\, t\hskip-0.5mm=\hskip-0.5mm0.
  $ }
\label{fig:7}\end{figure}

Another type of a large wave may be obtained by substituting
$$ \alpha=1, \;\;   \rho \to i\rho , \; \;\;\;\chi \to \chi+i\displaystyle\frac{\pi}{2}   , \; \;\;\;\gamma \to \gamma+i\displaystyle\frac{\pi}{4}   \eqno{\rm(3.14)}$$
into (3.11), which gives us
$$
\hskip-3mm \begin{array}{lll}
&f(x,y,t)=2\displaystyle\frac{\partial^2}{\partial x^2}\ln  \Big[2\lambda\Upsilon +  {\cosh 2\Gamma } \Big],& \\\\
&\Upsilon \hskip-1mm=\hskip-1mm\rho \hskip-1mm+\hskip-1mm x\sinh \chi\hskip-1mm  -\hskip-1mm2\Big[  {\lambda\cosh  \chi } \hskip-1mm+\hskip-1mm\mu\sinh  \chi  \Big]y\hskip-1mm -\hskip-1mm12\Big[\lambda^2\sinh \chi \hskip-1mm+\hskip-1mm \mu^2\sinh  \chi \hskip-1mm +\hskip-1mm2 \lambda\mu\cosh  \chi \Big]t,
 &    \\\\
&\Gamma=  \gamma+\lambda x-2\lambda\mu y-4\lambda(\lambda^2+3 \mu^2)t.
&
\end{array} \hskip-4mm {\rm(3.15)}
$$
  where $\chi, \lambda, \mu,   \rho $ are real constants,  $  \gamma  $ is either a real constant or a real constant $\pm \displaystyle\frac \pi 2 i$. Regularization (3.7) of (3.15) is shown in Figure \ref{fig:7}.

Due to (2.43) the fluid moves to the right whenever $f>0,$ such regions are shown in Figure \ref{fig:7}  in light shades; the fluid moves  to the left whenever $f<0$, such regions are shown in Figure  \ref{fig:7} in dark shades. When the fluid moving to the right collides with the fluid moving to the left, we obtain a "collision"  of light and dark lines. The velocity $\pmb v=(v_x, v_y)$ of motion of the "collision" is the same as the velocity of the motion of the wave profile, its components  may be uniquely determined from   the system of equations
$$\begin{array}{l}
   v_x  -2\Big[ {\lambda   \coth  \chi } +\mu   \Big]v_y= 12\Big[ \lambda^2  + \mu^2  +2  \lambda\mu   \coth\chi \Big]  \\ \\
  v_x-2 \mu v_y= 4 (\lambda^2+3 \mu^2)
\end{array}\eqno{\rm (3.16a)}$$
obtained from (3.13a)  by means of substitutions (3.14). Since the determinant of system (3.16a) is always nonzero,   the system has
a unique solution
$$
v_x=4\lambda^2-12\mu^2- {8\lambda\mu}{\tanh \chi},\quad   v_y=- {4\lambda}{\tanh \chi}-12\mu.
\eqno{\rm (3.16b)}
$$
The velocity is finite for all values of $\chi.$
 The velocity $\pmb v=(v_x, v_y)$ is not to be confused with the velocity of the flow given by the vector $\pmb u^\prime,$ the former is the velocity of motion of the profile while the latter is the velocity of motion of the fluid.

\begin{figure}
\center{ {\includegraphics[scale=.4]{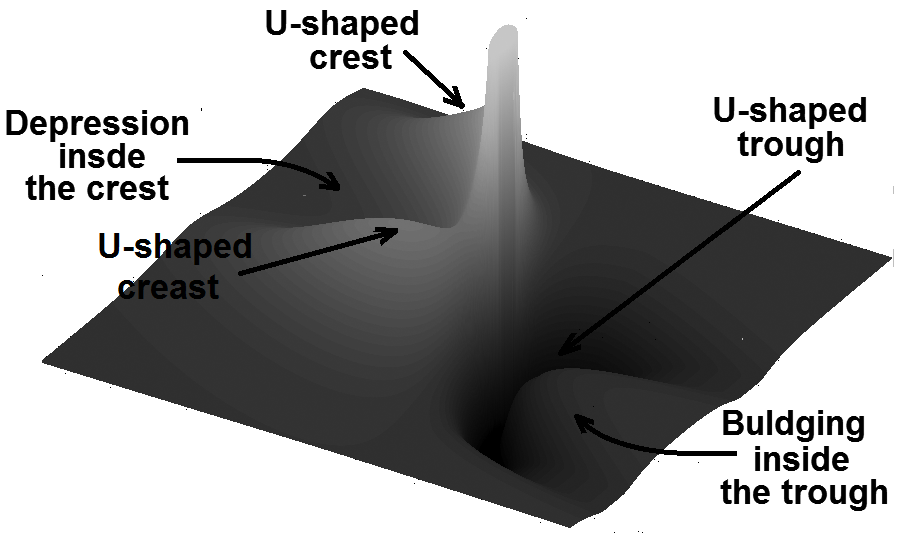}}}% Images in 100% size
  \caption{Central portion of the wave shown in Figure  \ref{fig:7} most likely to be observed with    terminology used in the text. \hskip90cm }
\label{fig:7aab}\end{figure}

\begin{figure}
\center{ {\includegraphics[scale=.58]{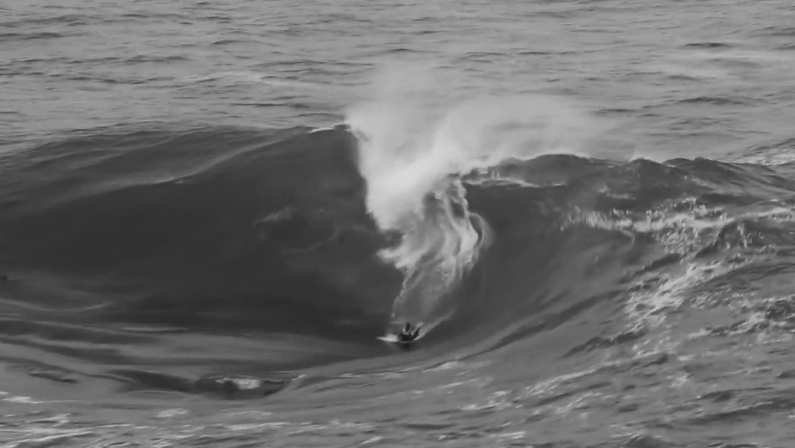}}}% Images in 100% size
  \caption{Physical wave  with the shape   predicted by Figure \ref{fig:7aab}. U-shaped crest, depression inside the crest and U-shaped trough are well-pronounced;  bulging inside the trough is seen but not well-pronounced. The picture is a frame from a video  shot  near Kiama on the NSW South Coast. Appeared  in \cite{Uwave1}.  }
\label{fig:Uwave1}\end{figure}

\begin{figure}
\center{ {\includegraphics[scale=.58]{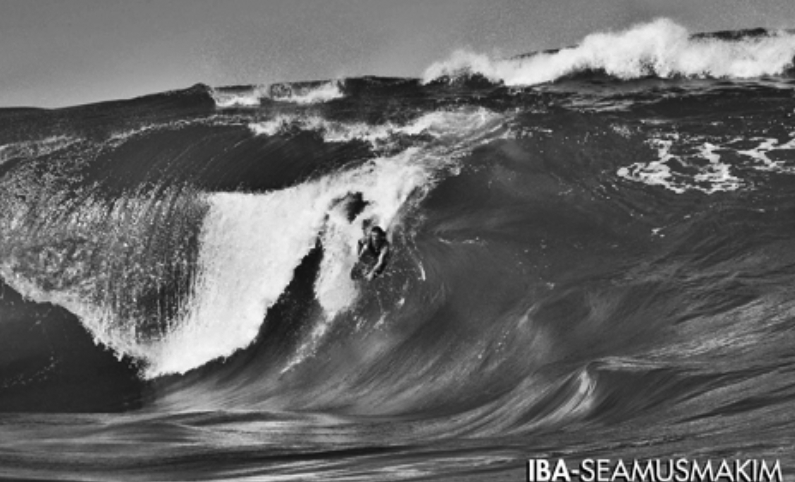}}}% Images in 100% size
  \caption{Physical wave  with the shape   predicted by Figure \ref{fig:7aab}. U-shaped trough and bulging inside the trough  are seen; U-shaped crest and depression inside the crest are barely seen.  The original photo was taken by   Seamus Makim near El Fronton, Canary Islands in 2010.  }
\label{fig:Uwave2}\end{figure}

\begin{figure}
\center{ {\includegraphics[scale=.58]{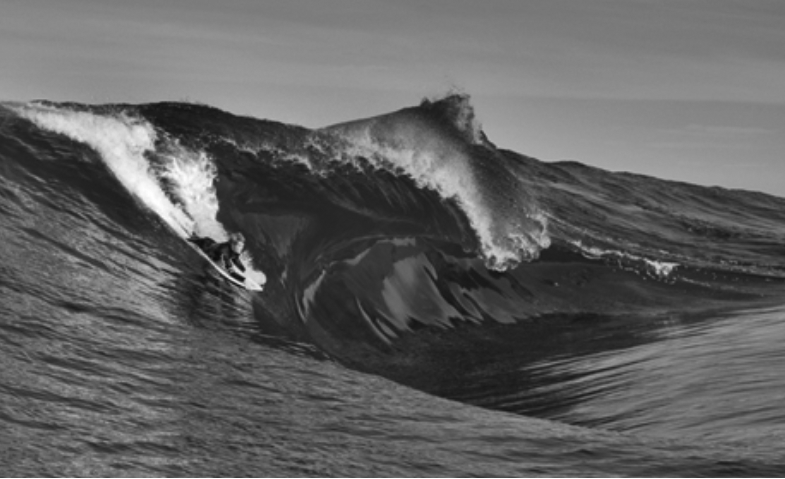}}}% Images in 100% size
  \caption{Physical wave  with the shape  predicted by Figure \ref{fig:7aab}. U-shaped trough and bulging inside the trough  are well-pronounced; U-shaped crest and depression inside the crest are not seen and may nor may not be present.  The original photo was taken by   Mike Maxted  in south-west Western Australia. Appeared  in \cite{Uwave4}.}
\label{fig:Uwave4}\end{figure}

 Waves of type (3.15) would be extremely unstable, so it would be very unlikely to see one as shown in Figure   \ref{fig:7}.  However the portions shown in Figure \ref{fig:7aab} should be observable at least for short periods. Indeed, Figures \ref{fig:Uwave1}-\ref{fig:Uwave4} show such waves.

%Permission was granted by  Mitch Coslovich
%  in email received May 18 2012 at 11:08pm
%Permission was granted by  Seamus Makim
%  in email received May 18 2012 at 10:19pm
%Permission was granted by Mike Maxted
%  in email received May 18 2012 at 6:19pm

Formula (2.43) implies that the higher up are the points of the fluid the faster they move in the  $x$ direction ultimately leading to wave overturning which destroys  the wave.  However, neither  solutions (3.11) nor solutions  (3.15), (3.18)    exhibit wave overturning due to the limitations of KP equations. This  limits the applicability and usefulness of  solutions (3.11)-(3.15) yet they do describe certain aspects of the corresponding physical waves such as sharp crests, the presence of troughs, the motion as a whole, etc.
 The  life-span of the corresponding physical waves might be relatively long or fairly short depending on how long conditions (2.1)-(2.9) are sustained.

The scaling transformation
$$
x \to x \delta,\hskip2mm y \to  y {\delta^2}, \hskip2mm t \to   t {\delta^3}, \hskip2mm f \to \displaystyle\frac f  {\delta^3} \eqno{\rm (3.17b)}
$$
equivalent to
$$
\lambda \to \displaystyle\frac \lambda \delta,\hskip2mm \mu \to \displaystyle\frac \mu \delta,   \eqno{\rm (3.17a)}
$$
transforms each of the solutions of KP considered  above  into  a solution of KP of the same type, the velocity undergoes transformation
$$
v_x\to \displaystyle\frac  {v_x}{ \delta^2},\quad v_y \to \displaystyle\frac {v_y}\delta. \eqno{\rm (3.17c)}
$$
Transformation (3.17)  allows us to re-scale  the graph of the corresponding solution.

\section{  The true rogue waves.    }

\begin{figure}
  \centerline{\includegraphics[scale=.33]{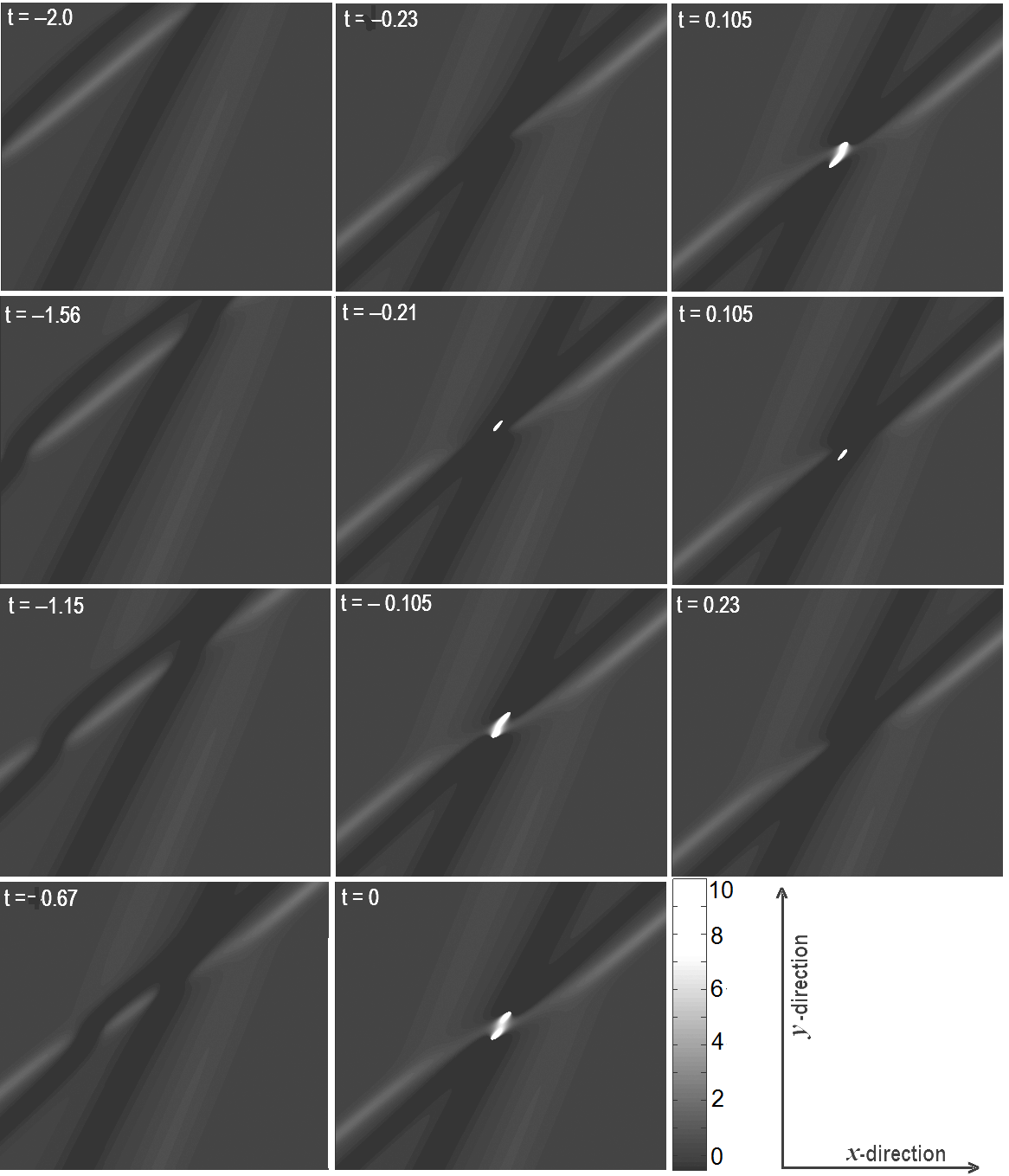}}% Images in 100% size
  \caption{Regularization  (4.1) with $M\hskip-1mm=\hskip-1mm10$  of time evolution of
(3.12) with
$N\hskip-1mm=\hskip-1mm2, \, \alpha\hskip-1mm=\hskip-1mm1, $ $  \chi_1\hskip-1mm=  0.6,\,
\lambda_1\hskip-1mm= 0.5,  \,
\mu_1 \hskip-1mm= 0.2,\,
\gamma_1\hskip-1mm=  0,\,
\rho_1\hskip-1mm= 0,\,
\chi_2\hskip-1mm=\hskip-1mm - 0.7 ,\,
\lambda_2\hskip-1mm= 1  ,\,
\mu_2\hskip-1mm=0.5 ,\,
\gamma_2\hskip-1mm=  0,\,
\rho_2\hskip-1mm= 0   $  in the region
 $-15< x,y<15 $. The  picture can be re-scaled using (3.17). }
\label{fig:8}\end{figure}

 In  the previous section we discussed the use of some singular solutions of KP to model large waves.  They may also be  used to provide a possible explanation  for true rogue waves.  Indeed, consider superposition of two waves of type (3.11) given by formulas (3.12).  Each wave moves in a corresponding direction with its "crossing" traveling laterally along the wave as discussed after formula  (3.13b).  When the "crossings" of the two waves confluence they produce a   drastic short-term jump  in amplitude localized in a small region much the same way true rogue waves do.    We refer to  such a  jump as { OTIN }which is an acronym for {  \underline{O}ne \underline{T}ime \underline{IN}tense  big wave/splash/wall of water/etc.}   The phenomenon is illustrated in Figure  \ref{fig:8} where for better visualization we plot ad-hoc re-normalization
 $$  f_M(x,y,t)=\left\{\begin{array}{l} \displaystyle\frac{f}{\vert f\vert }\ln \Big[\ln \Big(\Big \vert e^f-1 \Big\vert+1\Big)+1\Big], \textmd{ if } f\leq 0,\\\\ f, \textmd{ if } 0< f<M, \\\\
 M, \textmd{ if } f\geq M \gg1,\end{array} \right.,  \eqno{\rm (4.1)}$$
 rather than $f(x,y,t).$   Re-normalization  (4.1) completely cuts off the values of $f(x,y,t)$ above $M$ and provides a better resolution in  the range $0<f<M $.  The previously-defined  re-normalization  (3.7) does not cut off any values and allows us to look at all of $f(x,y,t)>0$ by using a logarithmic scale yet the very use of the logarithmic scale reduces the resolution. Just like (3.7) re-normalization (4.2) allows us   to   remove  physically unattainable large positive and negative values of $f(x,y,t)$ given  by (3.12).
 The height of the surface stays relatively low for $ t< -0.23$ and $t>0.23$ while  at the time interval $-0.23< t  <0.23$ a powerful  OTIN rises from the water seemingly out of nowhere only to  disappears seemingly into oblivion, the amplitude of the OTIN is at least three times the largest amplitude of the waves for $\vert t \vert > 0.23$. The confluence of the "crossings"  may create a very large  OTIN  even if the  two waves  themselves have   rather modest  amplitudes, making it appear as if the OTIN rises out of nowhere.

\begin{figure}
  \centerline{\includegraphics[scale=.38]{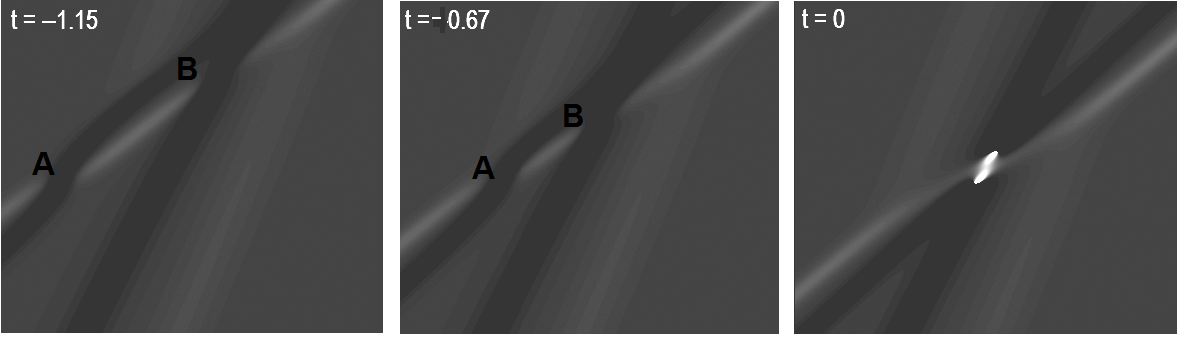}}% Images in 100% size
  \caption{The particles of water in
the white regions move to the right while at the dark regions they move to the left thus
creating undertows at points A and B pumping water from the lower layers to the crest.
At time t = 0 points A and B merge into one and so do the undertows leading a relatively
strong but very short-lived undertow pumping water from lower layers to the top and
creating a relatively huge but short-lived and spatially localized rogue wave  }
\label{fig:8a}\end{figure}

  To understand what is happening during the confluence of "crossings"  let us look at Figure  \ref{fig:8a} depicting three time frames  from   Figure  \ref{fig:8}.  According to (2.43)-(2.44) the   water in the white regions moves to the right and up while at the dark regions it moves to the left and up thus creating   undertows at points A and B  pumping water from the trough to the crest. At time $t=0$ points A and B merge into one while the two undertows intersect, at the intersection the water has no way to go other than up  creating a relatively large but short-lived and spatially localized rogue wave.  The formation of an OTIN is somewhat similar to the formation of a rip current.  Note that while an OTIN is rising the water moves up very fast resulting in a drop in water pressure under the OTIN, thus a pressure sensor placed placed under the OTIN would show a pressure drop rather than pressure hike.

\begin{figure}
  \centerline{\includegraphics[scale=.25]{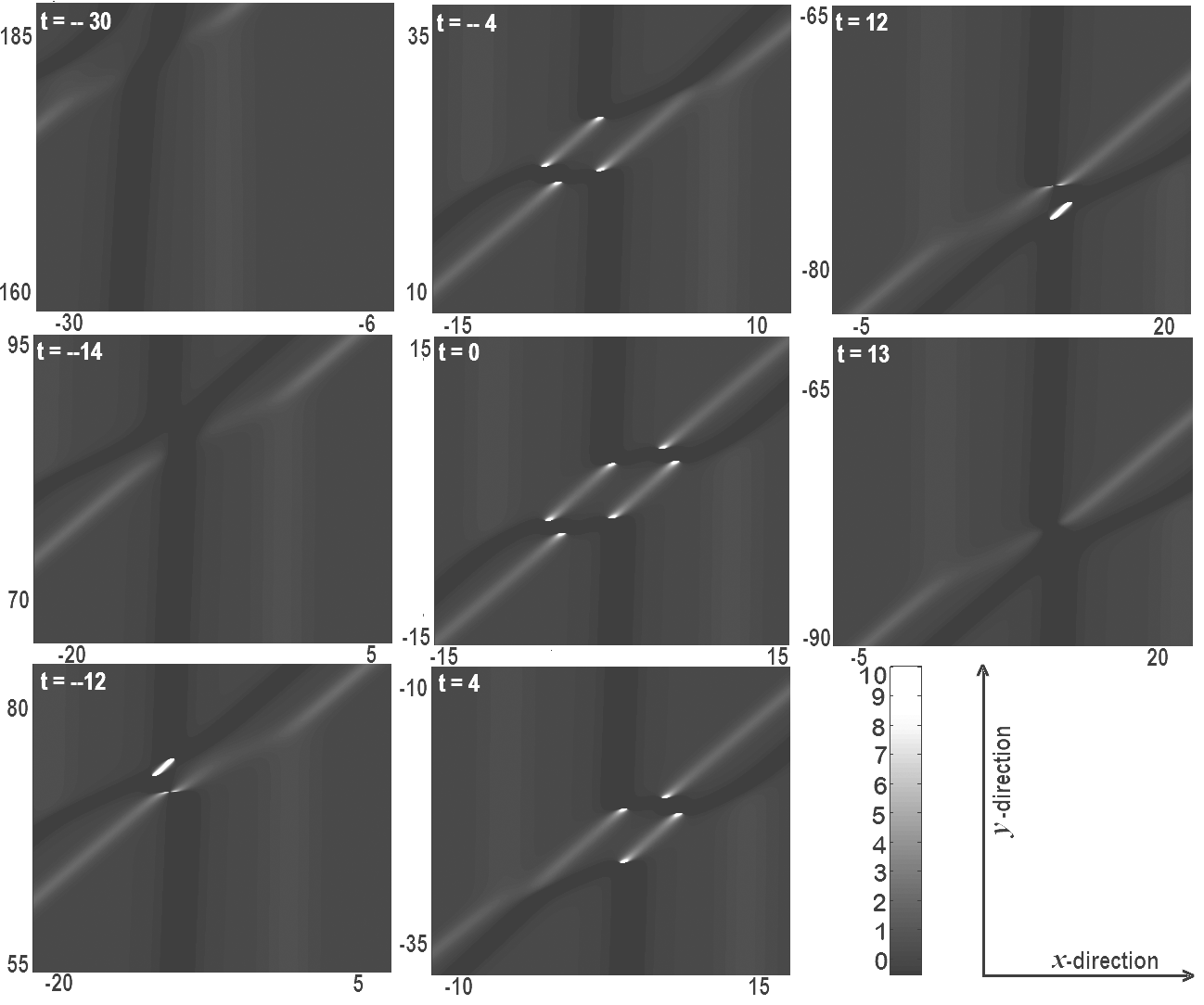}}% Images in 100% size
  \caption{ Regularization  (4.1) with $M=1$  of time evolution of
(3.12) with
$N=2, \, \alpha\hskip-1mm=\hskip-1mm1,\,
  \chi_1\hskip-1mm=\hskip-1mm 0.6,\, \lambda_1\hskip-1mm=\hskip-1mm0.5,  \,
\mu_1 \hskip-1mm=\hskip-1mm0.01,\,
\gamma_1\hskip-1mm=\hskip-1mm 0,\,
\rho_1\hskip-1mm=\hskip-1mm0,\,
\chi_2\hskip-1mm=\hskip-1mm- 0.7 ,\,
\lambda_2\hskip-1mm=\hskip-1mm1  ,\,
\mu_2\hskip-1mm=\hskip-1mm0.5 ,\,
\gamma_2\hskip-1mm=\hskip-1mm 0,\,
\rho_2\hskip-1mm=\hskip-1mm0
  $ }
\label{fig:9}\end{figure}

Under certain conditions an OTIN may split into several parts. To demonstrate it consider   wave  (3.12) with
$N=2, \, \alpha\hskip-1mm=\hskip-1mm1,\,  \chi_1\hskip-1mm=\hskip-1mm 0.6,\,
\lambda_1\hskip-1mm=\hskip-1mm0.5,  \,
\mu_1 \hskip-1mm=\hskip-1mm0.01,\,
\gamma_1\hskip-1mm=\hskip-1mm 0,\,
\rho_1\hskip-1mm=\hskip-1mm0,\,
\chi_2\hskip-1mm=\hskip-1mm- 0.7 ,\,
\lambda_2\hskip-1mm=\hskip-1mm1  ,\,
\mu_2\hskip-1mm=\hskip-1mm0.5 ,\,
\gamma_2\hskip-1mm=\hskip-1mm 0,\,
\rho_2\hskip-1mm=\hskip-1mm0  $, which differers from the previous example solely by the value of $\mu_1.$  Time evolution of re-normalization (4.1) of this function   is shown in Figure  \ref{fig:9} in a moving coordinate frame.
 Notice that the  OTIN appears to be  the largest and the most powerful  in the beginning before it breaks up and in the end after all the parts are recombined together. The event may appear to the crew of  a vessel carried by the current as a very powerful wave followed by   a few less powerful ones with another powerful wave int he end.   Whereas the first one hits the vessel which is still intact and with the relatively fresh crew it may not appear that strong,   the last one    hits the vessel already heavily pounded    with an exhausted crew. It is the last one that would appear to be  the most devastating to the crew thus leading to the legends of the "devyatiy val".

Figure  \ref{fig:9} also shows that   the small white spots indicating high elevation  cluster in groups of threes, e. g. the lower three white spots in the frame $t=-4$ , the upper  three white spots in the frame $t=4$,   the  upper three white spots   the lower three white spots in the frame $t=0$.  Some rogues waves have been reported to appear in groups of threes, e. g.  on April 16, 2005  a cruise liner Norwegian Dawn  encountered a series of three  21.34 m  rogue waves off the coast of Georgia;  in March 2012 a cruise liner Louis Majesty encountered   a series  of three 7.9 m rogue waves  between Cartagena and Marseille; there are reports of rogues waves appearing in groups of three on  Lake Superior.

OTINs rise in the regions shown in Figures  \ref{fig:8} -  \ref{fig:9}  in black or close to black shades, these are the  regions whose surface is considerably below the sea level and where a lot of underwater traffic takes place,  such regions are just giant troughs in the ocean.   Since the surface of such a trough is   below the sea level, it makes even a modestly tall OTIN to appear considerably taller, e.g. an 4 meter OTIN rising from a trough  3 meters deep appears to the crew of a ship that happens to be in the same trough to be 7 meters tall thus contributing to more drastic perception of the OTIN.

\begin{figure}
  \centerline{\includegraphics[scale=.33]{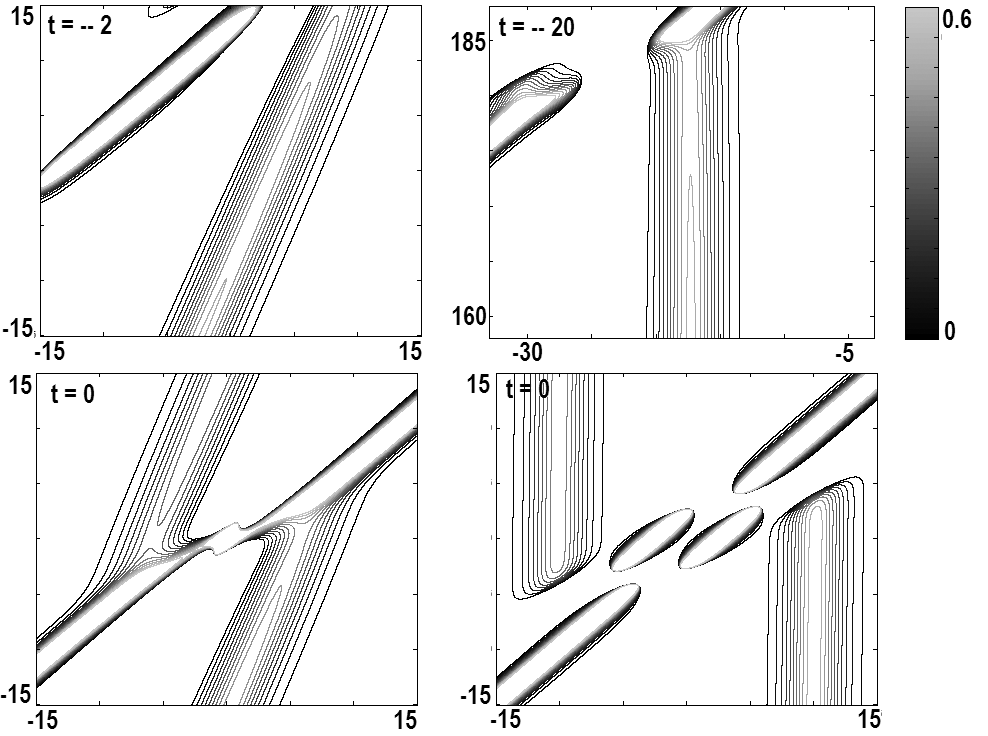}}% Images in 100% size
  \caption{ Contour maps of the portions $0< f<0.6 $of two frame of Figure \ref{fig:8} (in the first
      row) and two frame of Figure   \ref{fig:9}   (in the second row).  \hskip100mm  }
\label{fig:10}\end{figure}

Both graphs have portions somewhat resembling linear waves. Figure  \ref{fig:10}  provides contour maps of the portions $0< f<0.6 $ of two frames of Figure  \ref{fig:8}  (in the first row) and two frames of Figure  \ref{fig:9}   (in the second row).  These waves appear before, after and around the OTINs contributing to  the  appearance of calm waters and exacerbate the impression of OTINs appearing seemingly out of nowhere.

Figures   \ref{fig:8} - \ref{fig:9} illustrate OTINs that arise at the interaction of waves of the type pictured in Figure \ref{fig:4plus}.  When the waves pictured in   Figure \ref{fig:5plus} interact the result is not much different.  Transformation (3.17) with sufficiently large $\delta$   allows us to re-scale the functions  in Figure  \ref{fig:8}- \ref{fig:9}  to reduce the values of $f$ before and after the OTIN to very small while making the width of the crests and troughs very large;   physical waves corresponding to such functions would be practically invisible before and after the OTIN.

\section{   Is KP a good model for large and rogue waves?   }

In this paper we used singular solutions of KP to explain certain observed phenomena,   our description and explanation of  large and rogue waves  considerably differ  from other existing models.      So the natural question is "Is it valid?"  There are several points to consider.

The first one is whether we can use KP to describe waves in non-shallow water. The Kadomtsev-Petviashvili equation is typically derived for shallow water waves with condition (2.4) interpreted as the plane $z^{\prime}=-h$ being the bed of the body of water.  In this paper we assume that  the body of  water is not necessarily shallow not necessarily shallow but
the relevant flow is restricted to a near-surface layer   $z^{\prime}\ge -h$  below which the vertical component of the flow can be neglected. Such an  assumption may not bode well with many a researcher who are used to thinking that any motion on the surface should be felt all the way down to the very bottom, yet anyone with even a minimal SCUBA diving experience can attest to the fact that often even very powerful waves on the surface of the ocean have very little if any at all effect at the depths of even 10-15 meters.   What contributes to the disappearance  of the flow at greater  depths  is the viscosity of water small but nevertheless nonzero, internal flows, the motion of plankton and fish, interaction with waves reflected from the floor, etc.   In view of all of this, our assumption that the relevant portion of the (not necessarily shallow)  flow  is restricted to a rather thin
near-surface layer   $z^{\prime}\ge -h$ with (2.4) on $z^{\prime}= -h$   is not unreasonable.

\begin{figure}
  \centerline{\includegraphics[scale=.22]{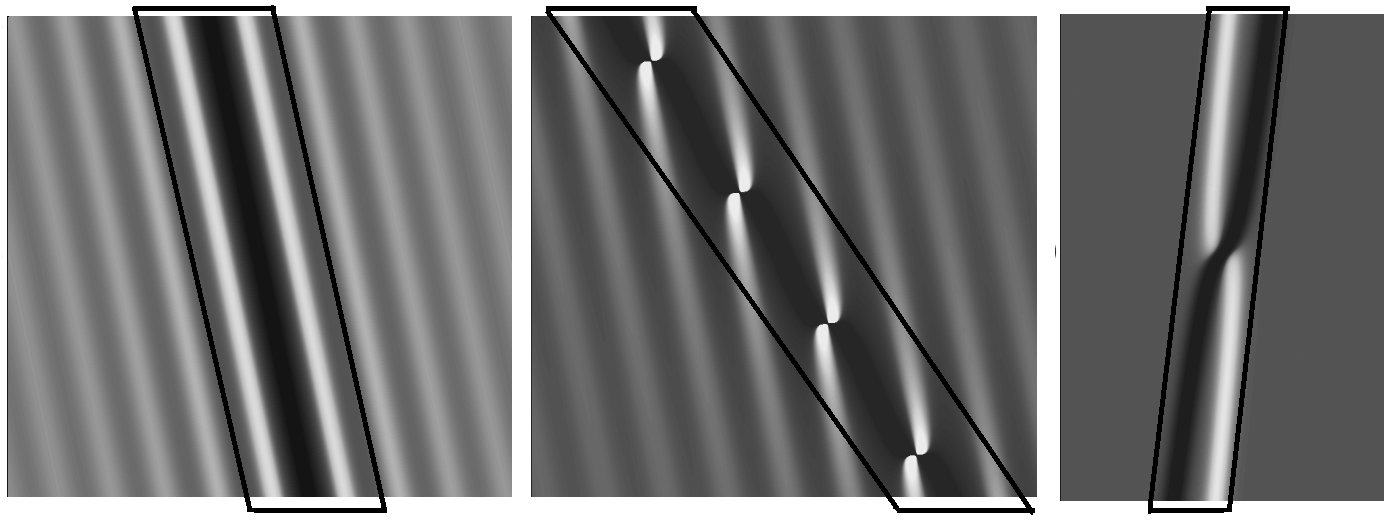}}% Images in 100% size
  \caption{ These are the  solutions shown in Figures  \ref{fig:1}, \ref{fig:3},  \ref{fig:4plus}.   The regions outside black parallelograms  are within the regime of KP while the regions inside the black parallelograms are not. \hskip150mm  }
\label{fig:singular}\end{figure}

The second point is whether we may use KP to describe waves of relatively large amplitude. Indeed, why would solutions of an equation derived for waves of relatively small amplitude work for waves of a relatively large amplitude?  If we look closer at the functions  considered such as the ones reproduced in   Figure  \ref{fig:singular} we will see that each such function consists of two parts: one is outside the black parallelogram and the other one is inside. The former is of small amplitude and very well within the regime of KP, it is the latter that contains a crest of large amplitude and a trough extending all the way to $-\infty $  taking these  functions outside of the regime  of KP.
However, since the  largest portion of the solution shown is outside of the black parallelogram and  within the regime of KP it must describe the corresponding physical wave quite well; we may assume therefore that its continuation inside the parallelogram is at least qualitatively  similar to the physical wave.  Singular solutions are often discarded as nonphysical which may be the case for some, but as shown in this paper the singular solutions discussed  in the paper provide a rather good qualitative description of corresponding physical waves.  The very existence of the singular curve is somewhat contradictory as we assume the flow is shallow yet the singularity extends all the way down to $-\infty$. Estimate   (2.43) allows us  to interpret the singularity as an indication of a  strong undertow in the region near the singularity, yet since KP is not capable of describing such an undertow the singularities appear.  The physical waves corresponding to such solutions do not have crests as high as predicted by the solutions and troughs extending all the way to $-\infty$, but their crests are still relatively  tall and troughs are relatively deep. The   solutions of KP discussed in this paper are often   referred to as "mathematical caricatures" rather than "mathematical models" due to the fact that just like regular caricatures they provide rather sketchy description of the phenomena with some features exaggerated and some others    suppressed.

\begin{figure}
  \centerline{\includegraphics[scale=.2]{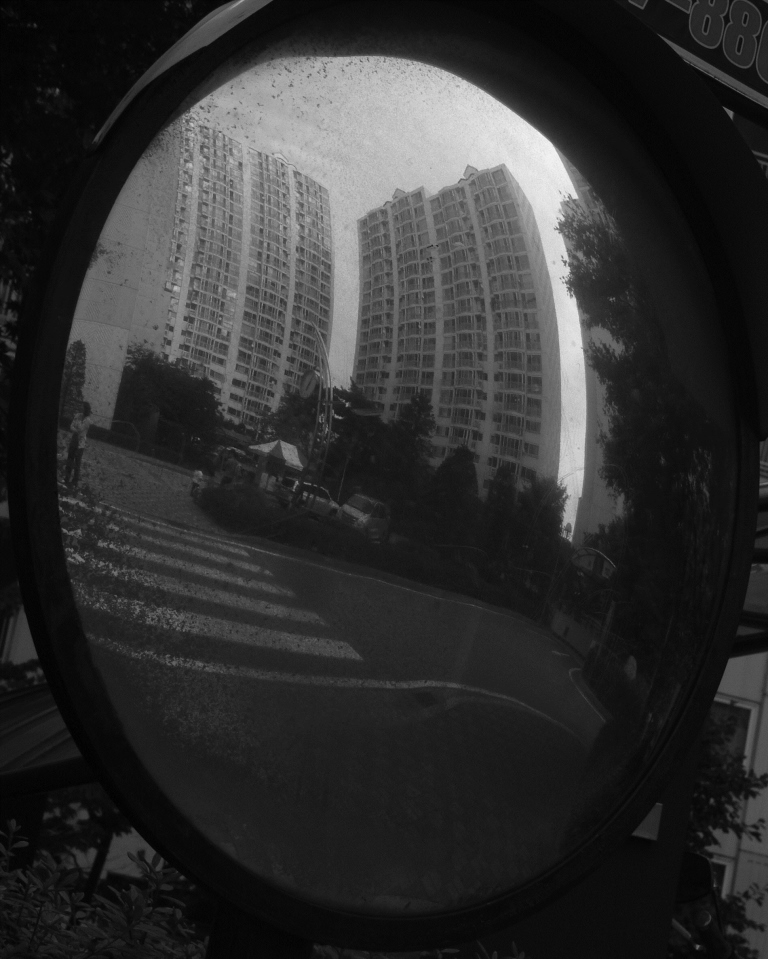}}% Images in 100% size
  \caption{  A typical curved mirror found everywhere around the world at road intersections.   A small mosquito flying a few centimeters away from the mirror would see an almost perfect reflection of itself but the image of a large building would be significantly distorted.  Yet even distorted,  the building's features could be easily seen and distinguished.  \hskip100mm  }
\label{fig:mirror}\end{figure}

Philosophically speaking we may compare  the Kadomtsev-Petviashvili equation  to a  curved mirror shown in Figure  \ref{fig:mirror}.  A small mosquito flying a few centimeters away from the mirror would see an almost perfect reflection of itself but the image of a large building would be significantly distorted.  Yet even distorted,  the building's features could be easily seen and distinguished.  Just like the curved mirror the  Kadomtsev-Petviashvili equation produces almost a perfect reflection of certain small waves with its soliton solutions;  larger waves are described by singular solutions which distort the dimensions but manage to preserve  main features of the waves  providing a unified albeit very primitive qualitative theory of large and rogue waves predicting    the existence of "crossings" for the waves shown in Figure   \ref{fig:4plus}, the existence of bulge inside the trough for the waves shown in Figure   \ref{fig:7aab}, the existence of OTINs, the clustering of rogues waves in groups of threes as shown in Figure \ref{fig:9},   the appearance of rogue waves seemingly out of nowhere and after a short while disappearance  seemingly  without a trace, etc. In other words, they show that there is "a method to the mess" and the same underlying principles are responsible for all these phenomena.

We should also notice that the solutions of KP discussed do not take into account wave overturning/breaking    and behave as if wave  overturning/breaking does not exist even though formula (2.43) predicts wave overturning/breaking as it implies that the higher points on the crest move faster than the lower ones. To incorporate overturning/breaking  into the model KP needs to be replaced with another equation, unfortunately what kind of equation the author does not know yet.

 Although it might be tempting to replace the singular solutions discussed in the paper with regular ones  expressible in  terms of  the Riemann $\theta-$function with appropriately chosen parameters, such a replacement might  not be very useful.  Such a replacement would allow  us to remove the singularities at the price of replacing rather simple solutions  with much more complicated ones which have essentially the same behavior.   The very presence of singular curves allows us to separate wave (3.5)  into halves given  by (3.9) or the waves shown in Figures \ref{fig:4plus}, \ref{fig:4minus} into waves shown in Figures \ref{fig:5plus}, \ref{fig:5minus}.

     In \S 3 singular solutions and their superpositions are given by formulas (3.1), (3.2), (3.5), (3.6), (3.11), (3.12), (3.15),   yet so far the author has not succeeded in combining   these formulas  into a single one. The singular solutions discussed here are only the simplest singular solutions of  KP,  it is possible and very likely that   singular solutions not discussed here  may shed some light on other physical phenomena.

For completeness sake we mention that there have been previous attempts to model rogue waves using KP.  For example,  \cite{gb, pp} tried to model rogue waves using two-soliton solutions. Their main idea is predicated
  on the observation that the value of such a solution is considerably larger at the intersection of the solitons and thus may represent a rogue wave. Such  attempts, however,  were doomed to fail from the start  mainly because physical rogue waves appear seemingly out of nowhere and after a short while disappear seemingly without a trace but the two-soliton solutions preserve their shape and move as a whole with velocity given by (3.4).    There have been numerous attempts to explain rogue waves in terms of modulation  instability for KP, formation of $\delta-$function type of solutions of KP, etc. as well as some properties of solutions of the Korteweg-de Vries equation, the nonlinear Schrodinger equation and several other equations.

  \section*{   Acknowledgments.  }

The author would like to thank Professor Hubert Chanson for kindly providing his picture of an undular bore; Roger Brooke and Andres Grawin  for permission to reproduce the picture of the tsunami of December 26, 2004;  Scott McClimont of Rip Curl and the authors for permission to use the frames from the video "Tip 2 Tip - Seven Ghosts  The Teaser" ;   Dr. Susanne Lehner and    Deutsches Zentrum fur Luft- und Raumfahrt for permission  to reproduce Figure \ref{fig:w17}; Shane Ackerman and Mitch Coslovich for permission to use Figure \ref{fig:Uwave1};   Mike Maxted  for permission to use Figure \ref{fig:Uwave4};  Seamus Makim for permission to use Figure \ref{fig:Uwave1}; Tim Lesson and the staff of the Riptide online magazine for help with Figures  \ref{fig:Uwave1}-\ref{fig:Uwave4}.   The original pictures were trimmed and turned into black and white by the author of this paper who takes the responsibility for any loss in technical or artistic quality of the pictures.

Although only the work directly related to this paper is referenced and listed in Bibliography, there has been a lot of research in the areas of  both singular solutions of integrable equations and rogue waves. The author would like to acknowledge all the work done in both fields and not mentioned here; the references are omitted not out of disrespect but merely due to the policy of more and more journals requiring that only the directly related work be referenced.

Part of the calculations presented in this paper was verified by Mr.  Wayne in his MS thesis in the Department of Mathematics of the University of Alberta under the supervision of the author and  co-supervision of Dr. Bica.

During the work on the paper the author interviewed numerous fishermen and boatmen on the islands of Cebu and Mactan in the Republic of the Philippines who witnessed or experienced large and rogue waves first hand. The author would like to express his gratitude to all of them even though their names the author has not kept.

% You may incorporate your references as follows in your main tex file.
% Using BibTex is not recommended but can be handled.

\medskip
% The data information below will be filled by AIMS editorial staff
Received xxxx 20xx; revised xxxx 20xx.
\medskip

\end{document}